# Hidden interfacial magnetism in $RuO_2$ heterostructures revealed by a superconducting spin switch

Tetsuma Mandokoro[1], Hisakazu Matsuki[1,2*], Shutaro Karube[1,2], Rodrigo de las Heras[3], Andrei Mazanik[3], Akihiro Yamada[1], Fugo Tokoro[1], Susumu Hirata[1], Ryusuke Hisatomi[1,2], Yoichi Shiota[1,2], Vitaly N. Golovach[3,4,5,6], F. Sebastian Bergeret[3,4], and Teruo Ono[1,2,7*]

1. Institute for Chemical Research, Kyoto University, Uji, Kyoto 611-0011, Japan
2. Center for Spintronics Research Network, Institute for Chemical Research, Kyoto University, Uji, Kyoto 611-0011, Japan
3. Centro de Física de Materiales (CFM-MPC) Centro Mixto CSIC-UPV/EHU, E-20018 Donostia-San Sebastián, Spain
4. Donostia International Physics Center (DIPC), 20018 Donostia-San Sebastián, Spain
5. Departamento de Polímeros y Materiales Avanzados: Física, Química y Tecnología, Facultad de Química, University of the Basque Country UPV/EHU, 20080 Donostia-San Sebastián, Spain
6. IKERBASQUE, Basque Foundation for Science, 48013 Bilbao, Spain
7. International Center for Synchrotron Radiation Innovation Smart, Tohoku University, Sendai 980-8577, Japan

*E-mail: matsuki.hisakazu.2i@kyoto-u.ac.jp, and ono@scl.kyoto-u.ac.jp

**$RuO_2$, once the flagship altermagnetic candidate, is now believed to be nonmagnetic in the bulk, yet weak, static and spatially inhomogeneous magnetism emerges at its surfaces and interfaces. Where this magnetism resides, what stabilises it, and whether it is connected to these states remain open questions. Here we address these questions at the Nb/$RuO_2$ interface using a superconducting spin switch. A GdN/Nb/$RuO_2$ trilayer allows interfacial exchange-field alignment to tune the superconducting transition temperature of Nb, probing interfacial magnetism inaccessible to conventional magnetometry. The switch reveals a hidden magnetic subsystem, which we attribute to a thin $RuO_x$ redox layer formed as Nb scavenges oxygen from $RuO_2$. This layer behaves as a weak and magnetically hard ferromagnet with vacancy-induced anisotropy. Furthermore, supercurrent amplifies the switching efficiency, making the probe sensitivity tunable *in situ.* Superconducting spin switches thus emerge as a quantitative low-temperature probe of hidden magnetism at buried interfaces.**

Few materials have had as turbulent a magnetic biography as $RuO_2$. For decades, bulk $RuO_2$ was regarded as a Pauli paramagnetic metal[1,2], before evidence for itinerant antiferromagnetism emerged[3,4]. $RuO_2$ subsequently became the prototypical *d*-wave altermagnet, predicted to exhibit anisotropic momentum-dependent spin splitting whose magnitude and sign depend on crystal orientation and Néel vector direction despite the absence of net magnetisation[5]. This picture was supported by resonant X-ray[6], photoemission[7] and transport signatures[8–11]. However, it has been progressively challenged by null results, particularly in the bulk: muon spin rotation (μSR)[12,13] and neutron diffraction[14] find no static magnetic order, while angle-resolved photoemission[15,16] and quantum oscillations[17,18] reveal a Fermi surface consistent with a nonmagnetic electronic structure. Thin-films studies have likewise reported no magnetic order[19] and transport signatures consistent with a non-altermagnetic origin[20], while similar spin-torque responses in nonmagnetic $IrO_2$ undermined the torque-based evidence for altermagnetism[21]. However, magnetism has not disappeared entirely. Low-energy μSR on thin-films reveals weak, static, and spatially inhomogeneous magnetism localised near surfaces[22], strained films exhibit enhanced magnetic responses[23], and interfacial or defect-related moments continue to emerge in heterostructures[24-26]. Intriguingly, uniquely-strained $RuO_2$ films also exhibit superconductivity[27–29], underscoring how profoundly the electronic ground state can be reconstructed away from the bulk limit. The emerging picture is therefore that magnetism in $RuO_2$ is not an intrinsic bulk band property, but an emergent local phenomenon highly sensitive to surfaces, strains, interfaces and stoichiometry[30]. What has been missing is an experimental probe capable of detecting and quantifying such weak magnetism and a microscopic mechanism explaining its most puzzling combination of properties: a moment too small for standard magnetometry, and magnetic order hard enough to survive as a stable, switchable reference.

Here we supply both. We use a superconducting spin switch in which a superconductor is sandwiched between two magnetic layers ($M_1$/S/$M_2$)[31]. When the superconducting thickness is less than its coherence length $\xi_0$, the interfacial exchange fields from $M_1$ and $M_2$ add or subtract for parallel (P) or antiparallel (AP) magnetisations, giving distinct superconducting transition temperatures $T_c$(P) and $T_c$(AP), with $\Delta T_c = T_c(\mathrm{AP}) - T_c(\mathrm{P})$[32–40]. Comparing $T_c$ between these configurations provides a sensitive differential probe of interfacial exchange fields difficult to isolate in S/M bilayers, particularly in the presence of magnetic domains[41–43]. Using GdN/Nb/$RuO_2$(110) trilayers, we observe reversible $T_c$ switching independent of the crystallographic direction of field-cooling (FC) and stable against applied fields of several hundred millitesla. These observations are inconsistent with a bulk altermagnetic origin of the switching. Control experiments localise the magnetic subsystem responsible for the switching at the Nb/$RuO_2$ interface. We propose that its magnetic layer originates from a $RuO_x$ redox layer that necessarily forms as Nb scavenges oxygen from $RuO_2$. Theoretical analysis of the redox chemistry suggests that this redox layer is a thin ferromagnet with a small magnetic moment. The resulting weak Zeeman coupling implies that large fields are required for magnetisation reversal. The local anisotropy axes are randomly oriented hard axes, as in amorphous magnets[44], because the oxygen vacancies that create the moments also determine their spin–orbit anisotropy. Their preferential alignment towards the interface normal produces a coherent easy plane, keeping the moments in-plane and making the switching insensitive to crystallographic axes. Meanwhile, the random component is expected to pin the order, and to favour its breakup into in-plane domains beyond a characteristic length $L^*$ because random anisotropy defeats exchange on large scale[45,46]. This layer is therefore weak in moment and hard in orientation, invisible in magnetometry yet stable enough to serve as a magnetic reference for the switch. The magnetisation of the $RuO_x$ layer induces an effective exchange field at the Nb/$RuO_x$ interface. We model the superconducting spin switch with the quasiclassical Usadel equation[47–49], which allows us to extract the effective interfacial exchange field, and accounts for the increase of $|\Delta T_c|$ with applied supercurrent. Together with zero-field-cooling (ZFC) measurements, this establishes the spin switch as a sensitive transport probe of buried interfacial magnetic order.

## Superconducting spin switch

A AlN(10 nm)/GdN(10 nm)/Nb(9.3 nm)/$RuO_2$(110)(10 nm)//$TiO_2$(110) heterostructure was deposited by magnetron sputtering as shown in Fig. 1**a** (see **Methods**). The epitaxial growth of $RuO_2$ was confirmed by reflection high-energy electron diffraction (RHEED) *in-situ* and X-ray diffraction (XRD) as shown in Figs. 1**b, c**. The thin-film was patterned into four-terminal bars of 10 μm × 220 μm for electrical measurements as shown in Fig. 1**d**. Figure 1**e** shows the magnetisation versus in-plane magnetic field, $M(H)$, of the unpatterned film at 10 K. GdN is a ferromagnetic insulator with a Curie temperature of ~40 K and an effective switching field of ~6 mT at 10 K (see Supplementary S1), consistent with previous reports[39,50]. Its magnetisation can therefore be reversed with millitesla fields while leaving the harder magnetic subsystem unchanged (see Supplementary S1 about $M(T)$ and $M(H)$ at different temperatures). No discernible difference was observed between the magnetisation curves after ±7 T FC from room temperature and ZFC, indicating that conventional magnetometry does not resolve the magnetic subsystem investigated below (see Supplementary S2 for the discussion about the magnetisation of Nb/$RuO_2$ bilayers). In some literature, the current and FC directions are aligned along the $RuO_2$ [001] axis, which has been proposed as the Néel vector direction in altermagnetic $RuO_2$[51,52]. In a d-wave altermagnet, the anisotropic exchange field is expected to produce an angular dependence of $\Delta T_c$ on the magnetisation direction at the opposite interface[53]. Here, however, [001] specifies only the experimental geometry and does not imply any magnetic significance. As we show below, the magnetically active subsystem resides in a $RuO_x$ redox layer at the Nb/$RuO_2$ interface and therefore does not inherit the hypothetical Néel vector orientation of the rutile $RuO_2$ lattice.

The switch is operated as follows. The device is field-cooled from room temperature in ±9 T along [001], setting the hard magnetic subsystem as shown in Figs. 2**a,b,d,e**; alternatively, it is zero-field-cooled after oscillatory demagnetisation at room temperature in Figs. 2**c,f**. At a few kelvins, the GdN magnetisation is saturated at a preset-field $\mu_0 H_{set}$=±50 mT along [001] and the field is returned to zero, defining the "±0 mT" states. The superconducting transition is then recorded upon warming at a current density of $5.38 \times 10^8$ A/m$^2$, with the current applied along [001]. After FC at +9 T, $T_c$ splits between GdN magnetisation along the positive field (along the FC direction, P) and negative field (opposite to the FC direction, AP) directions, giving $\Delta T_c = T_c(AP) - T_c(P) = -8$ mK as shown in Fig. 2**d**. Reversing the FC direction to –9 T reverses the $T_c$ switching with respect to the external field direction while preserving the relative response, $\Delta T_c$ = –8 mK in Fig. 2**e**. After ZFC, by contrast, $\Delta T_c$ vanishes as shown in Fig. 2**f**. Similar switching is observed with $\mu_0 H$=±50 mT (see Supplementary S3).

To identify the magnetic switching field associated with the $RuO_x$ redox layer, we measured the field dependence of the resistance of the AlN(10 nm)/GdN(10 nm)/Nb(9.3 nm)/$RuO_2$(110)(10 nm)//$TiO_2$(110) sample measured at 3.1 K, within the superconducting transition, as shown in Fig. 3**a**. Apart from a background associated with the suppression of superconductivity by the magnetic field, pronounced hysteretic anomalies appear around ±400 to ±500 mT. In contrast, no corresponding anomaly is observed at 5 K, where the Nb is in the normal state. The disappearance of these features above indicates that they reflect a magnetic modulation of superconductivity rather than ordinary normal-state magnetoresistance. Similar resistance peaks localized near the coercive field have been reported in ferromagnet/superconductor hybrids[42]. This assignment is independently supported by the remanent superconducting response shown in Fig. 3**b**. After field cooling at +9 T, the superconducting transition depends strongly on whether the sample is preset to -400 or -500 mT before returning to zero field. In other words, when the sample is preset to -400 mT, the sign of $\Delta T_c$ shift is consistent with that in Figs. 2**a** and 2**d**, whereas presetting to -500 mT reverses the sign, making it consistent with Figs. 2**b** and 2**e**.

Moreover, magnetic-field-dependent measurements of $T_c$ on the AlN(10 nm)/GdN(10 nm)/Nb(9.3 nm)//$SiO_2$/Si control sample revealed a positive sign for the exchange field induced by GdN within our model (see Supplementary S4), implying a negative sign for the exchange field induced by $RuO_x$. The sign of the exchange field $h_{\text{ex}}$ coincides with that of the imaginary part of the spin-mixing conductance, $G_i \simeq \pi G_0 N_F h_{\text{ex}} d_{\text{Nb}}$, where $G_0$ is the conductance quantum and $N_F$ is the density of states per spin at the Fermi level[40,54,55]. Unlike conventional superconducting spin switches with two equivalent superconductor/magnet interfaces, where the absolute sign of the exchange field cannot be determined, our asymmetric structure allows this sign to be inferred[31]. This negative exchange field is consistent with the itinerant picture in which the $t_{2g}$ band of $Ru^{3+}$ ($t_{2g}^5$) is more than half filled, so that Nb electrons hybridize with the unoccupied minority-spin states. The same sign follows in the atomic limit, as antiferromagnetic exchange between Nb conduction electrons and the $Ru^{3+}$ spins (see Supplementary S11).

The origin of the subsystem is further confirmed by a series of control experiments. (i) Crystallographic orientation: the switching measurements were repeated with current, cooling-field and preset-field applied along [$1\bar{1}0$] and along the in-plane direction midway between [001] and [$1\bar{1}0$], i.e. 45° from both axes, yielding comparable $\Delta T_c$ values (Supplementary S5). The switching is therefore insensitive to crystallographic orientation, arguing against an origin associated with the proposed altermagnetic $RuO_2$. (ii) Orthogonal geometry: with the cooling-field along [001] but the preset-field used to switch GdN rotated by 90°, $\Delta T_c$ vanishes, as expected for a superconducting switch controlled by the relative angle between two in-plane magnetisations[31,56] (Supplementary S6). (iii) Nb thickness dependence: increasing the Nb thickness to 12 nm, above $\xi_0$ (~11.35 nm), also suppresses $\Delta T_c$ to zero. The switching therefore requires the exchange perturbations from the two interfaces to overlap coherently across the Nb layer, as expected for a genuine two-interface exchange field effect[31,32] (see Supplementary S7 for $\xi_0$ and S8 for the switching of the 12-nm-thick Nb device). This thickness dependence rules out alternative mechanisms based on stray fields[57,58] or spin-polarised quasiparticles[59,60], which do not require $d_{Nb} < \xi_0$. (iv) No $RuO_2$: an AlN(10 nm)/GdN(10 nm)/Nb(9.3 nm)//$SiO_2$/Si control sample shows no dependence of $T_c$ on the GdN magnetisation direction and, decisively, the same ±9 T FC protocol produces no evidence of a hard magnetic subsystem. This excludes the GdN/Nb interface as its origin (see Supplementary S9). (v) No GdN: an AlN(10 nm)/Nb(9.3 nm)/$RuO_2$(110)(10 nm)//$TiO_2$(110) control sample shows no measurable $T_c$ response upon reversing the cooling-field direction (see Supplementary S9). (vi) $RuO_2$ thickness dependence: increasing the $RuO_2$ thickness to 50 nm, beyond the relevant proximity length in $RuO_2$, preserves the switching polarity observed in the 10-nm-thick $RuO_2$ sample (see Supplementary S10). For a 50-nm-thick $RuO_2$ layer, the $TiO_2$/$RuO_2$ interface is far beyond the relevant superconducting proximity length from the Nb layer. Therefore, Cooper pairs are not expected to reach the $TiO_2$/$RuO_2$ interface and the persistence of switching in the thicker $RuO_2$ sample supports an origin near the Nb/$RuO_2$ interface. Thus, the hard magnetic subsystem at the Nb/$RuO_2$ interface cannot be independently resolved through $T_c$ in the bilayer geometry; GdN provides the second, switchable exchange field required for its differential detection. Together, these controls localise the hard magnetic subsystem to the Nb/$RuO_2$ interface and establish the superconducting spin switch as its readout.

Together, these observations reveal an unusual combination of magnetism. The interfacial magnetic state retains the orientation set during FC and remains stable against external fields up to 400 mT. Despite this large coercive field scale, magnetometry of the unpatterned film stack detects no corresponding remanence and coercive field signal within experimental resolution (see Supplementary S2). The subsystem therefore combines a small net moment with hysteresis and rigidity. The origin of this weak and hard subsystem is addressed in the next section.

## A weak and magnetically hard magnet at the Nb/$RuO_2$ redox interface

The Nb/$RuO_2$ interface is not chemically inert. Per mole of oxygen, the formation enthalpies of the niobium oxides exceed that of $RuO_2$ by more than a factor of two[61–63], providing a strong thermodynamic drive for *in-situ*-deposited Nb to scavenge oxygen from the $RuO_2$ surface. Reduction of $Ru^{4+}$ to $Ru^{3+}$ at the Nb/$RuO_2$ interface, combined with oxygen diffusion at room temperature, therefore can produce a thin self-limited $Ru^{3+}$-rich layer at the interface. The oxygen vacancies play a dual role in generating magnetism. First, reduction creates $Ru^{3+}$ sites with one hole in the $t_{2g}$ multiplet ($t_{2g}^5$), which in the atomic limit forms a Kramers doublet with a robust $S = ½$ moment, while the loss of Ru-O bonds narrows the magnetically active planar *d*-band. The Stoner criterion, not satisfied in stoichiometric $RuO_2$, can therefore be reached in the reduced interfacial layer, which orders as an itinerant ferromagnet. Second, the same vacancies generate strong local symmetry breaking. Through spin–orbit coupling, each reduced Ru site acquires a magnetic hard axis associated with its local vacancy environment. These axes are random from site to site, but their ensemble is combed towards the interface normal by the vacancy-concentration gradient. The redox reaction therefore can create both the magnetic order and the random anisotropy governing its macroscopic behaviour.

The resulting state is a thin ferromagnet with random anisotropy, of the type studied in amorphous magnets[44–46]. The preferential alignment of the local hard axes towards the interface normal produces a coherent easy plane, whereas the random component competes with the exchange stiffness. In this picture, the energy gained by following the random axes grows faster with length scale than the exchange energy cost, leading to the predicted formation of in-plane domains beyond the Imry–Ma length $L^*$[45,46] (see Supplementary S11). Independently of the domain structure, the small magnetisation results in weak Zeeman coupling to an external field. Therefore, magnetisation reversal, whether by coherent rotation or by depinning of domain walls from the random-anisotropy landscape[64], requires large fields even for moderate anisotropy energy. The same oxygen vacancies that enable the Stoner instability through band narrowing also pin the resulting order through random anisotropy. This naturally produces the experimentally observed weak and rigid state—a magnet with an almost undetectable net moment, yet an orientation stable against large magnetic fields.

## Interfacial exchange fields and current-enhanced $\Delta T_c$

For $d_{Nb} < \xi_0$, one can assume that the superconducting order parameter is uniform across the Nb film, and the two magnetic interfaces can be represented by thickness-averaged effective exchange fields whose combination depends on the relative magnetisation alignment, $h_{\mathrm{eff,P(AP)}} = h_{\mathrm{GdN}} \pm h_{\mathrm{RuOx}}$. The parameter $h_{\mathrm{GdN/RuOx}}$ is the effective exchange field averaged over the superconducting layer, induced by each magnetic interface. The critical temperatures of the diffusive Nb film in the respective magnetic configurations, in the presence of a supercurrent, then follow from the self-consistently linearised Usadel theory[48,65,66]. Using $h_{\mathrm{GdN}} = 0.13$ meV estimated from Ref. 37, we obtain $h_{\mathrm{RuOx}}$ to be $-0.004$ meV (see Supplementary S12 about the calculation details). This highlights the superconducting switch as a sensitive probe of interfacial magnetism, capable of detecting an exchange field more than 30 times weaker than that of GdN.

We next examine the current dependence of $\Delta T_{\mathrm{c}}$. Increasing the applied current from 0.05 mA ($5.38 \times 10^8$ A/m$^2$) to 0.225 mA ($2.42 \times 10^9$ A/m$^2$) enhances $\Delta T_{\mathrm{c}}$ after FC ±9 T by approximately a factor of two as shown in Fig. 4**a**. In contrast, after ZFC, $\Delta T_{\mathrm{c}}$ remains small over entire current range. The quasiclassical model qualitatively reproduces this enhancement and reveals physics beyond the Ginzburg-Landau description. Expanding the digamma function to leading order gives additive contributions from current-induced depairing and exchange pair-breaking. The quasiclassical expression, however, retains higher-order cross terms between the two pair-breaking mechanisms, allowing depairing to enhance the sensitivity of $T_{\mathrm{c}}$ to the interfacial exchange field. Figure 4**b** shows the theoretical calculation of $\Delta T_{\mathrm{c}}$ depending on the $q\xi_0$. $q$ is expressed with respect to $\xi_0$ and characterises the depairing induced by the supercurrent. Here, $\xi_0 = \sqrt{\hbar D/\Delta_0}$ , where $D$ is the diffusion constant in Nb and $\Delta_0$ is the zero-temperature superconducting gap of the reference film in the absence of pair-breaking. The model reveals that increasing depairing enhances $|\Delta T_c|$. This behavior arises because $T_c^{\mathrm{P}} > T_c^{\mathrm{AP}}$, such that depairing suppresses $T_c^{\mathrm{AP}}$ more strongly than $T_c^{\mathrm{P}}$, thereby increasing $|\Delta T_c|$. Establishing a quantitative relation between $q$ and the total current is beyond the scope of this work, owing to the likely inhomogeneity of the current distribution along the $10\ \mu\mathrm{m}$ dimension of the superconducting region[67] (See Supplementary S12 about further details).

## Conclusions and outlook

In summary, we have used a superconducting spin switch to reveal hidden magnetism at the $Nb/RuO_2$ interface, which we attribute to a weak and magnetically hard ferromagnetic state in an interfacial $RuO_x$ redox layer. The switching provides a sensitive measure of the relative interfacial exchange fields, with its response amplified by supercurrent through the super-additive interplay between exchange fields and current-induced pair-breaking. The magnetic subsystem detected by the switch is not a bulk property of $RuO_2$, consistent with the null results of bulk probes[12,14,15,17], but is associated with an interfacial redox layer whose magnetism is both enabled and pinned by oxygen vacancies. This mechanism accounts for the seemingly contradictory properties that have made $RuO_2$ interfacial magnetism so elusive: a weak magnetic moment, undetectable by conventional magnetometry, yet static on μSR timescales[22], spatially inhomogeneous, and remarkably hard against magnetic-field reversal. The resulting picture is of a magnet that is weak in moment and hard in orientation. Although Nb promotes oxygen removal in the present heterostructure, a similar vacancy-driven mechanism is expected to operate in other oxygen-deficient $RuO_2$ systems, including commonly studied $RuO_2$/Py heterostructures where oxygen vacancies are present[25]. For magnetism, if its random-anisotropy character is confirmed, it may provide a solid-state realisation of a generalised Imry-Ma model whose phase diagram–including domain morphology, wall chirality, and glassiness, tunable by the redox depth–remains open. More broadly, the interfacial system realised here provides a largely unexplored setting in which random-anisotropy magnetism, interfacial spin–orbit coupling, and $s$-wave superconductivity coexist at atomic proximity.

The inverse switching polarity further identifies a negative $h_{\mathrm{ex}}\,(G_i)$ at the $Nb/RuO_x$ interface. Established spin switches based on two layers of the same magnets have primarily exploited their magnetic configurations. Here, combining a known magnetic reference with an interface of unknown magnetic character enables the switch to probe the sign and magnitude of interfacial exchange itself. For superconductivity, this establishes the switch, operated at finite current, as a quantitative probe of interfacial magnetism invisible to conventional magnetometry, with a sensitivity that is itself current-tunable. This enhanced response may extend the accessible range of weak interfacial magnetism beyond what can readily be resolved by heavy-metal-based spin Hall magnetoresistance. We expect this probe to be transferable to the growing family of compensated and weakly magnetic materials whose interfaces, rather than their bulk, host the magnetism.

The microscopic origin of the inverse superconducting switching polarity observed here remains an important subject for future research. Comparative studies using materials with established negative spin polarisation, such as FeCr[68,69], could help clarify the relationship between spin polarisation, interfacial exchange and switching polarity. Further experiments, including depth-resolved low-energy µSR and spatially resolved XMCD-PEEM, could characterise the depth distribution and lateral inhomogeneity of the interfacial magnetism. Comparing the resulting magnetic correlation lengths with predictions of the random-anisotropy model would test whether the layer indeed reaches the Imry–Ma regime.

## Methods

### Film growth and device fabrication

Thin-films are deposited onto pre-cleaned 10 mm × 10 mm $TiO_2$(110) substrates using an Eiko ES-250S rf magnetron sputtering with a base pressure better than $6 \times 10^{-6}$ Pa. The 10-nm-thick $RuO_2$(110) films are grown epitaxially on $TiO_2$(110) by reactive sputtering at a substrate temperature of 400°C, with Ar and $O_2$ partial pressure of 0.24 Pa and 0.06 Pa, respectively. Nb is deposited in Ar at a pressure of 0.5 Pa, while GdN is grown by reactive sputtering from a Gd target with Ar and $N_2$ partial pressures of 0.46 Pa and 0.04 Pa, respectively. Finally, a 10-nm-thick AlN capping layer is grown by reactive sputtering from an Al target with Ar and $N_2$ partial pressures of 0.46 Pa and 0.04 Pa, respectively, to protect the structure.

### Magnetic measurements

Magnetic moment versus magnetic field measurements are performed using a Quantum Design Magnetic Property Measurement System (MPMS) equipped with a vibrating sample magnetometer superconducting quantum interference device (SQUID). Magnetic fields of up to 7 T were applied using the system's superconducting magnet.

### Transport measurements

Low-temperature current-voltage $I(V)$ measurements are performed in a four-terminal electrical setup using Au wires ultrasonically bonded to 80-nm-thick Au contact pads deposited on the devices. Measurements above 1.8 K are carried out in a Quantum Design Physical Property Measurement System (PPMS) with an in-plane magnetic field. Measurements below 1.8 K are performed in the same system using the He-3 option. $T_c$ is defined at half the normal-state resistance unless otherwise noted.

## Competing interests

The authors declare no competing interests.

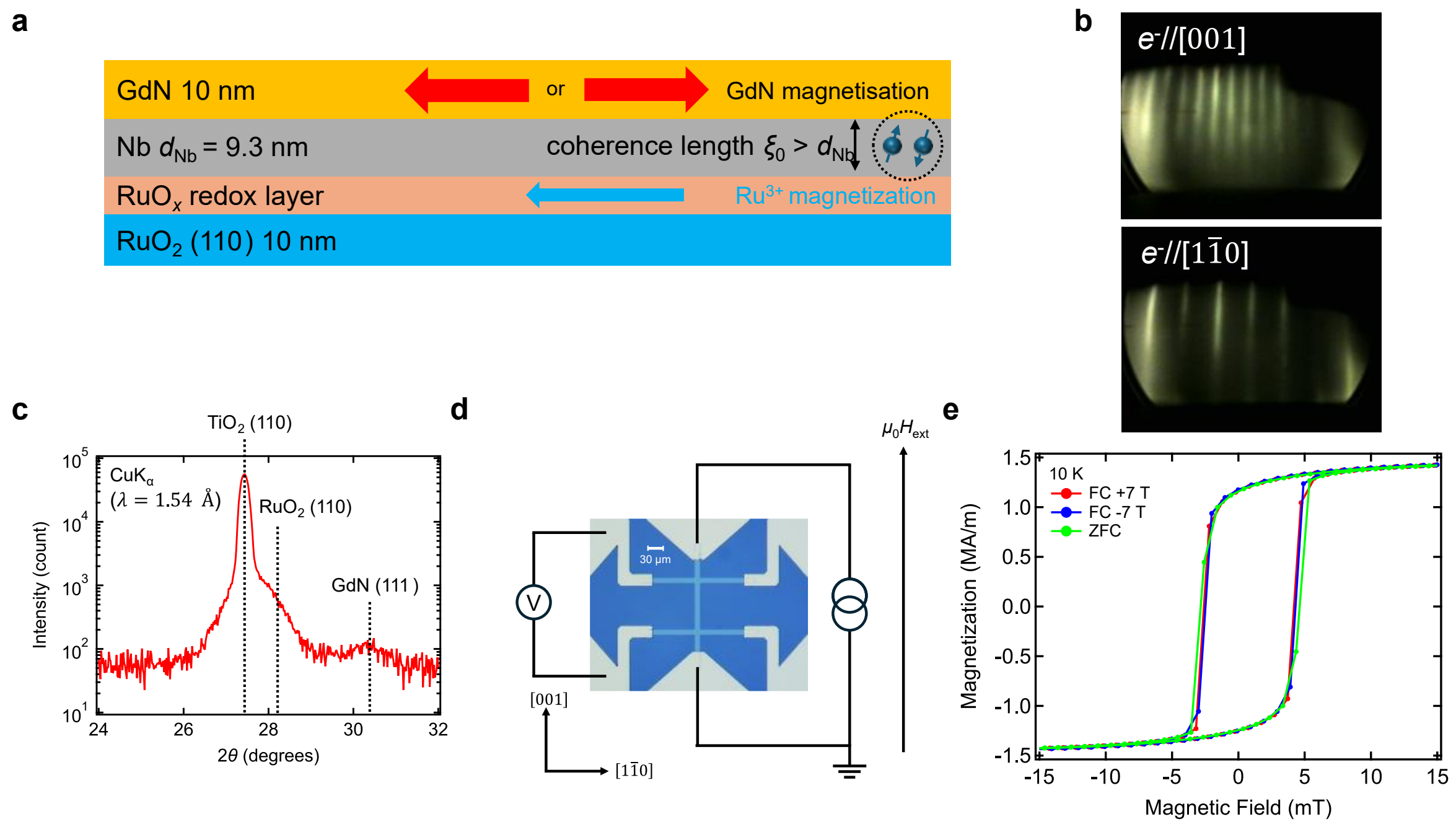

Fig. 1: **Structural, magnetic and device characterisation of the AlN(10 nm)/GdN(10 nm)/Nb(9.3 nm)/$RuO_2$(110)(10 nm)//$TiO_2$(110) superconducting spin switch. a**, Cross-sectional schematic of the AlN/GdN/Nb/$RuO_2$ heterostructure. The GdN layer and the $RuO_x$ redox layer provide the two magnetic subsystems, with their magnetisations indicated by the red and blue arrows,

respectively. Because the Nb thickness, 9.3 nm, is smaller than the superconducting coherence length, Cooper pairs in Nb can coherently experience the exchange fields from both the GdN/Nb and Nb/$RuO_x$ interfaces. **b**, RHEED image of the $RuO_2$(110) layer with the electron beam along the [001] and [$1\bar{1}0$] directions. **c**, XRD image of the heterostructure whereby the x-ray source is CuKα λ=1.54Å). **d**, Device structure and measurement configuration. **e**, Magnetisation as a function of in-plane magnetic field measured at 10 K, showing the magnetic hysteresis of the GdN layer. The magnetic field was applied along the [001] direction, which was also the field-cooling direction. Red, blue, and green solid lines show the magnetisation of the unprocessed film after field-cooling ±7 T and ZFC, respectively.

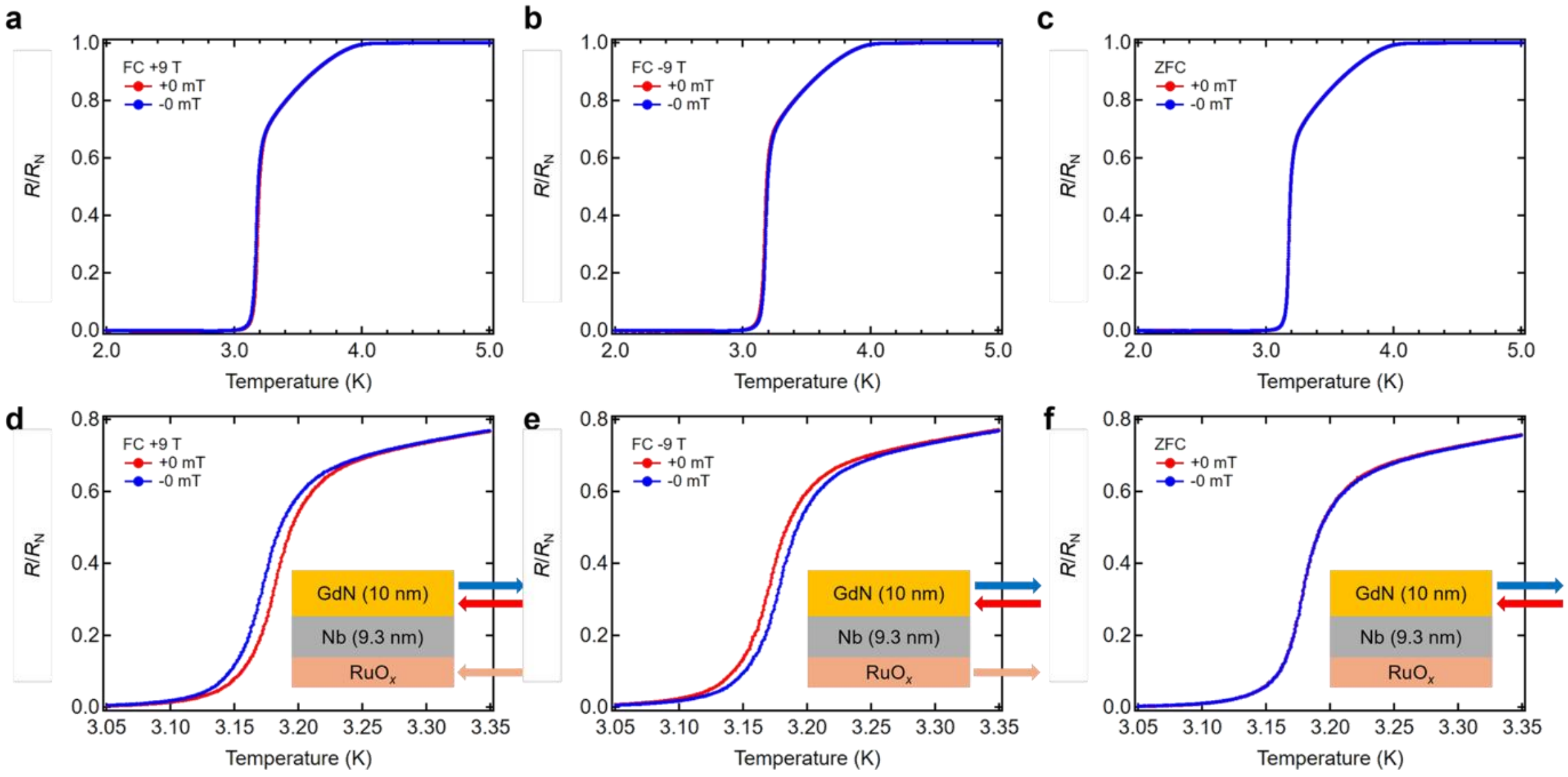

Fig. 2: ***$T_c$* of the switch for each magnetic configuration.** All measurements in zero applied field after saturating GdN at ±50 mT. **a-c,** Normalised *R*(*T*) after FC at +9 T, -9 T, and ZFC. **d-f,** Enlarged transition regions. After FC the two GdN polarities give distinct $T_c$, with sign set by the FC direction; after ZFC the difference vanishes. The insets sketch the relative orientation of the redox-layer magnetisation ***m*** (prepared by FC) and the GdN magnetisation.

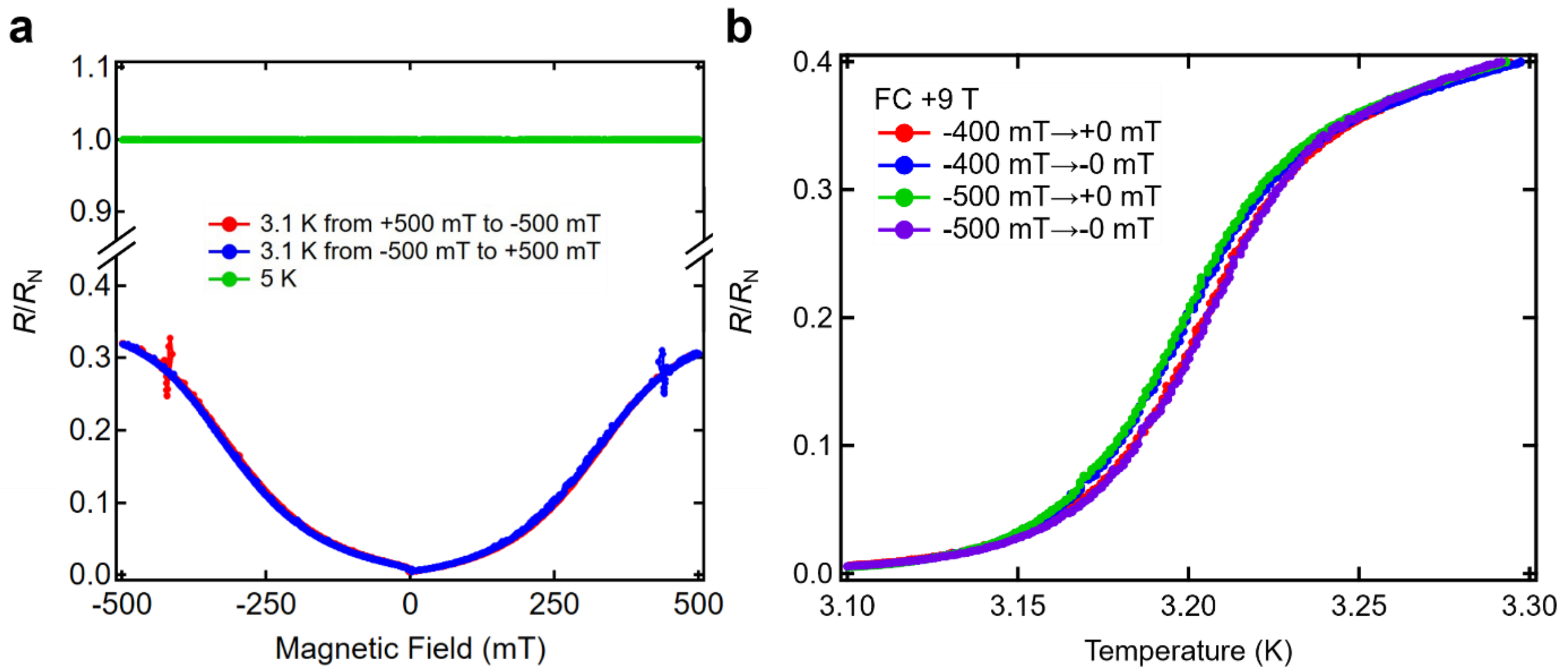

Fig. 3: **Switching field of the interfacial redox layer. a** Magnetic-field dependence of the resistance of the AlN(10 nm)/GdN(10 nm)/Nb(9.3 nm)/$RuO_2$(110)(10 nm)//$TiO_2$(110) sample measured at 3.1 K and at 5 K. The red and blue dots represent field sweeps from positive to negative field and from negative to positive field, respectively, at 3.1 K, while the green dots represent the field sweep measured at 5 K. **b** Zero-field superconducting transitions measured after field cooling at +9 T and subsequent presetting of the magnetic field. The red and blue curves correspond to measurements at +0 and -0 mT, respectively, after presetting to -400 mT, whereas the green and purple curves correspond to measurements at +0 and -0 mT, respectively, after presetting to -500 mT.

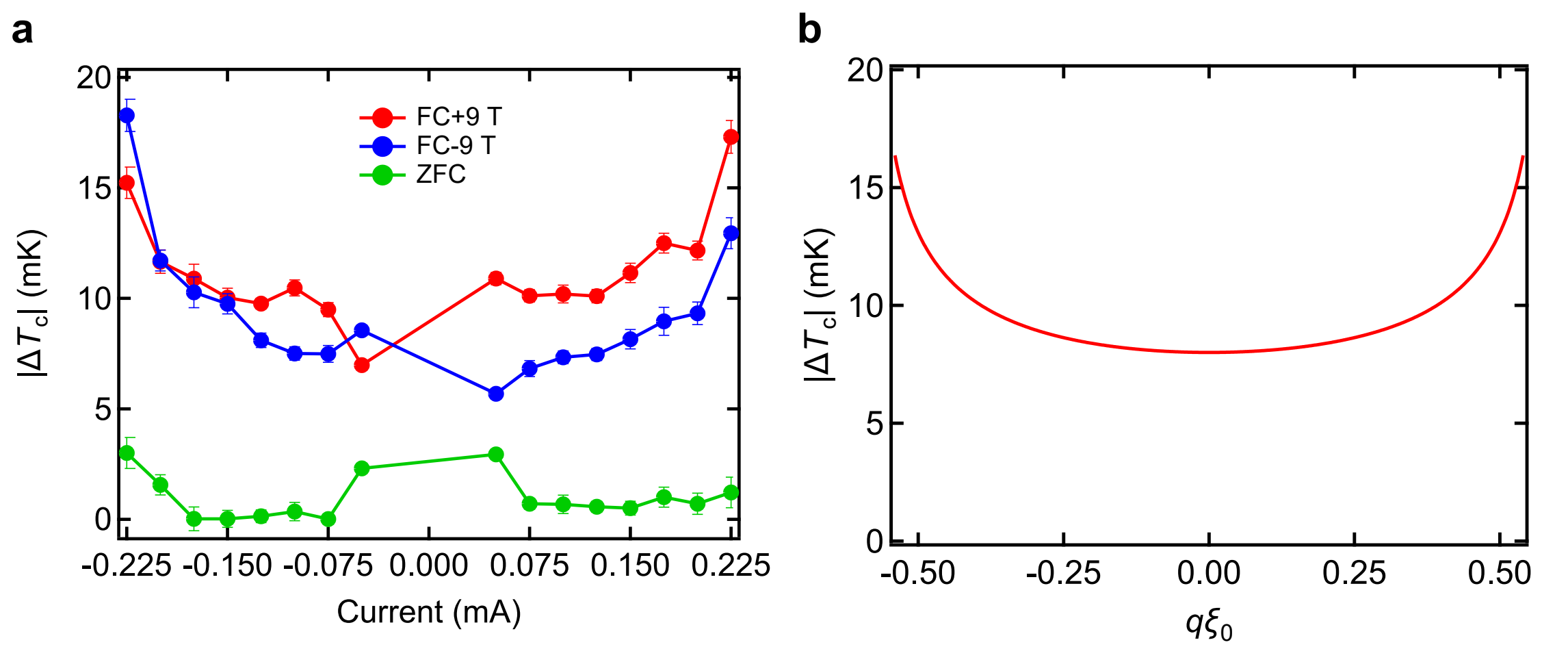

Fig. 4. **Supercurrent dependence of the difference of |$T_c$|. a,** Supercurrent dependence of $|\Delta T_c|$ under $\pm 0$ mT. The red, blue, and green dots represent the results obtained after FC at +9 T, -9 T, and ZFC, respectively. **b,** Calculated dependence of $|\Delta T_c|$ on $q\xi_0$, which characterises the

depairing induced by the supercurrent. The result was calculated using Eq. S5 with the estimated values of $h_{\mathrm{RuO}_x}$, $h_{\mathrm{GdN}}$, $\tau_{sf}$, and $\tau_{so}$.

# Supplementary Information

# Hidden interfacial magnetism in $RuO_2$ heterostructures revealed by a superconducting spin switch

Tetsuma Mandokoro,[1] Hisakazu Matsuki,[1,2,*] Shutaro Karube,[1,2] Rodrigo de las Heras,[3] Andrei Mazanik,[3] Akihiro Yamada,[1] Fugo Tokoro,[1] Susumu Hirata,[1] Ryusuke Hisatomi,[1,2] Yoichi Shiota,[1,2] Vitaly N. Golovach,[3,4,5,6] F. Sebastian Bergeret,[3,4] and Teruo Ono[1,2,7,†]

[1]*Institute for Chemical Research, Kyoto University, Uji, Kyoto 611-0011, Japan*

[2]*Center for Spintronics Research Network, Institute for Chemical Research, Kyoto University, Uji, Kyoto 611-0011, Japan*

[3]*Centro de Física de Materiales (CFM-MPC) Centro Mixto CSIC-UPV/EHU, E-20018 Donostia-San Sebastián, Spain*

[4]*Donostia International Physics Center (DIPC), 20018 Donostia-San Sebastián, Spain*

[5]*5. Departamento de Polímeros y Materiales Avanzados: Física, Química y Tecnología, Facultad de Química, University of the Basque Country UPV/EHU, 20080 Donostia-San Sebastián, Spain*

[6]*6. IKERBASQUE, Basque Foundation for Science, 48013 Bilbao, Spain*

[7]*International Center for Synchrotron Radiation Innovation Smart, Tohoku University, Sendai 980-8577, Japan*

[*] matsuki.hisakazu.2i@kyoto-u.ac.jp

[†] ono@scl.kyoto-u.ac.jp

## CONTENTS

## S1 Temperature dependence of the GdN magnetisation

Figure S1a shows the temperature dependence of the magnetisation for the GdN magnetisation in AlN(10 nm)/GdN(10 nm)/Nb(9.3 nm)/$RuO_2$(10 nm)//$TiO_2$(110) heterostructures. The Curie temperature of GdN is approximately 40 K. Figure S1b also shows the magnetic field dependence of magnetisation at 2 K, 4 K, and 6 K. Although the ferromagnetism of GdN strongly depends on the temperature as shown Fig. S1a, both the coercive field and remanence magnetisation remain nearly unchanged within the temperature range investigated in this study. These results are consistent with the properties of GdN reported in previous studies. [S1, 2].

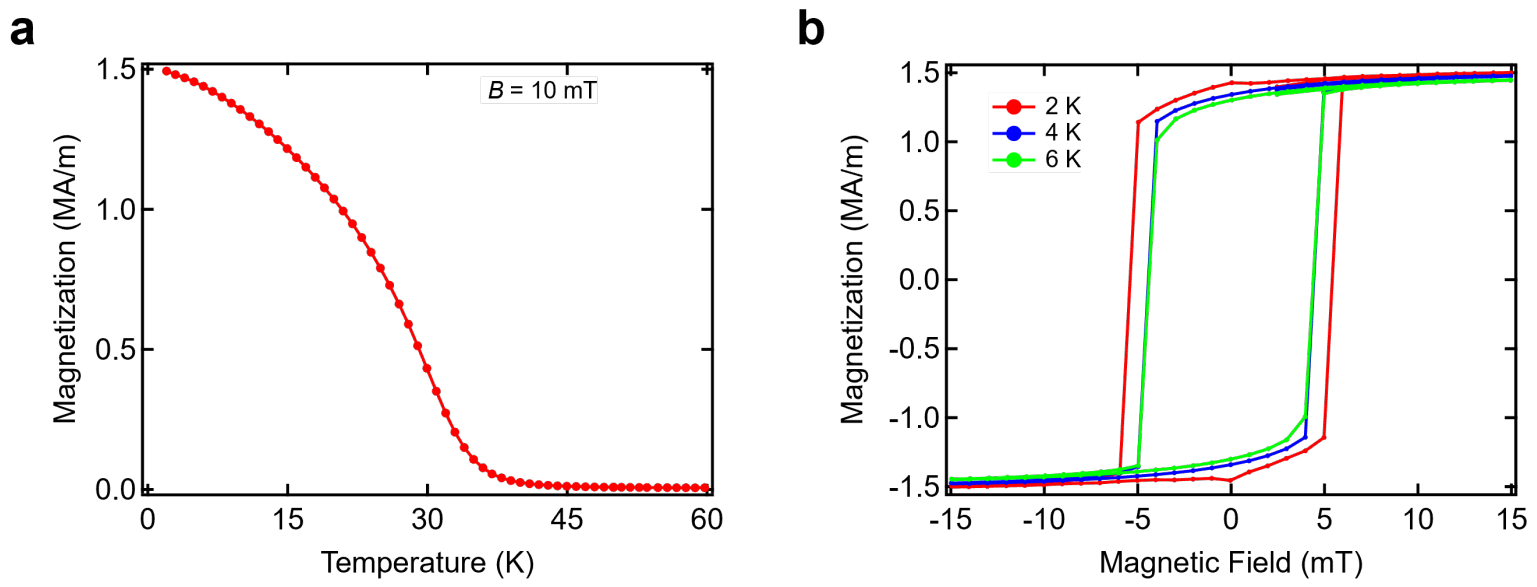

FIG. S1. (a) Temperature dependence of GdN magnetisation. (b) Magnetic field dependence of GdN. The red, blue, and green solid lines indicate the data measured at 2, 4, and 6 K, respectively.

**S2 Magnetometry of the AlN/Nb/$RuO_2$//$TiO_2$ and variability of the $TiO_2$ substrate-background**

To examine whether the magnetisation of $RuO_x$ redox layer inferred from the superconducting transport measurements is accompanied by a detectable macroscopic magnetic moment, we performed magnetic-field-dependent magnetisation measurements by SQUID. Figure S2a compares the magnetic responses of AlN(10 nm)/GdN(10 nm)/Nb(9.3 nm)/$RuO_2$(10 nm)//$TiO_2$ and AlN(10 nm)/Nb(9.3 nm)/$RuO_2$(10 nm)//$TiO_2$ samples. Although the contribution from the GdN layer is clearly visible in the AlN/GdN/Nb/$RuO_2$//$TiO_2$, almost no magnetic component attributable to the AlN/Nb/$RuO_2$//$TiO_2$ sample can be resolved. Figure S2b shows the magnetic-field dependence of the AlN/Nb/$RuO_2$//$TiO_2$ sample alone, again revealing no clear macroscopic hysteresis corresponding to the switching-field scale identified by superconducting transport. Here, the background linear components have been subtracted from the data. A quantitative extraction of only $RuO_x$ magnetic contribution is complicated by the magnetic background of the $TiO_2$ substrate. As shown in Fig. S2c, the $M - H$ response of the same $TiO_2$ substrate changes appreciably between measurements performed two days after baking and one week after baking. This thermal-history and time dependence prevents a reliable subtraction of the substrate contribution at the level required to isolate the weak magnetic signal from the $RuO_x$ redox layer. We therefore refrain from assigning an absolute magnetic moment or coercive field to the $RuO_x$ magnetic state from the SQUID measurements. Nevertheless, the absence of a coercivity hysteretic signal in the magnetometry indicates that the switching observed in superconducting transport is not associated with a readily detectable macroscopic ferromagnetic moment. The magnetic moment of the $RuO_x$ layer can remain below the sensitivity imposed by the substrate background even though its local magnetic configuration strongly affects the adjacent superconducting Nb layer.

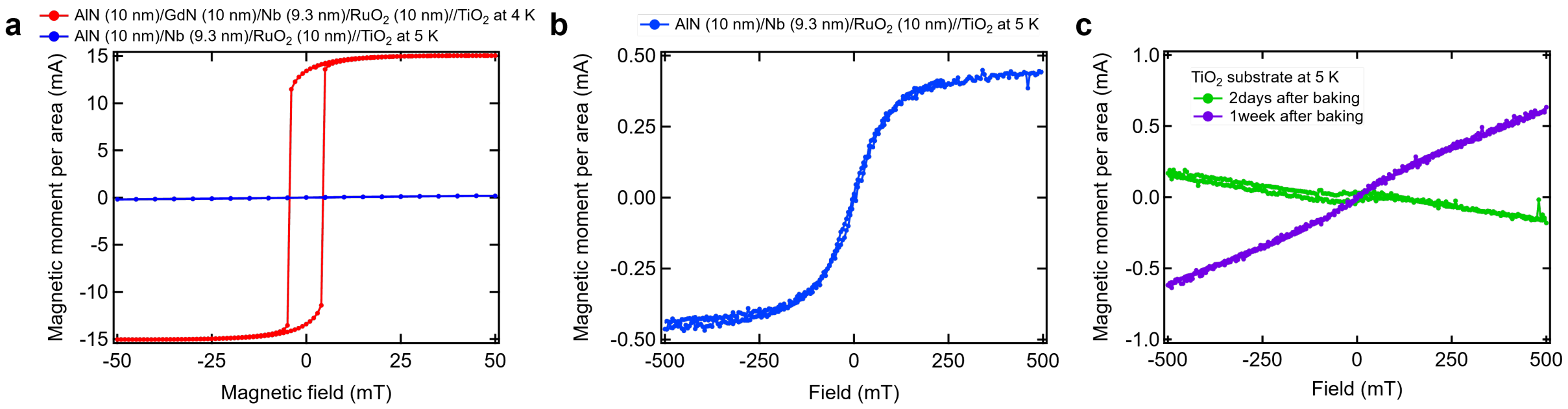

FIG. S2. (a) Magnetic-field dependence of the magnetic moment per area for the AlN/GdN/Nb/$RuO_2$//$TiO_2$ (red) and AlN/Nb/$RuO_2$//$TiO_2$ (blue) samples. (b) Enlarged magnetic-field dependence of the AlN/Nb/$RuO_2$//$TiO_2$ sample. (c) $M-H$ curves of the same $TiO_2$ substrate measured 2 days after baking (green) and one week after baking (purple).

**S3 Temperature dependence of the normalized Nb resistance under ±50 mT with current applied along [001]**

Figures S3a–c show the temperature dependence of the normalized resistance for the Nb 9.3-nm-thick sample measured with a current applied along the [001] direction under an external magnetic field of ±50 mT. Figures S3d–f show enlarged views of the superconducting transition region. Similar to the results obtained at ±0 mT in the main text, the superconducting transition temperature ($T_c$) changes depending on the relative orientation between the exchange field from $RuO_x$ and the magnetisation of GdN. Moreover, the sign of this shift is reversed by reversing the field-cooling (FC) direction. In contrast, no discernible difference in the transition temperature is observed after zero-field cooling (ZFC). Compared with the ±0 mT results, the difference in $T_c$ becomes smaller under the applied magnetic field presumably because the contribution from the $RuO_x$ exchange field is partially masked by the external magnetic field.

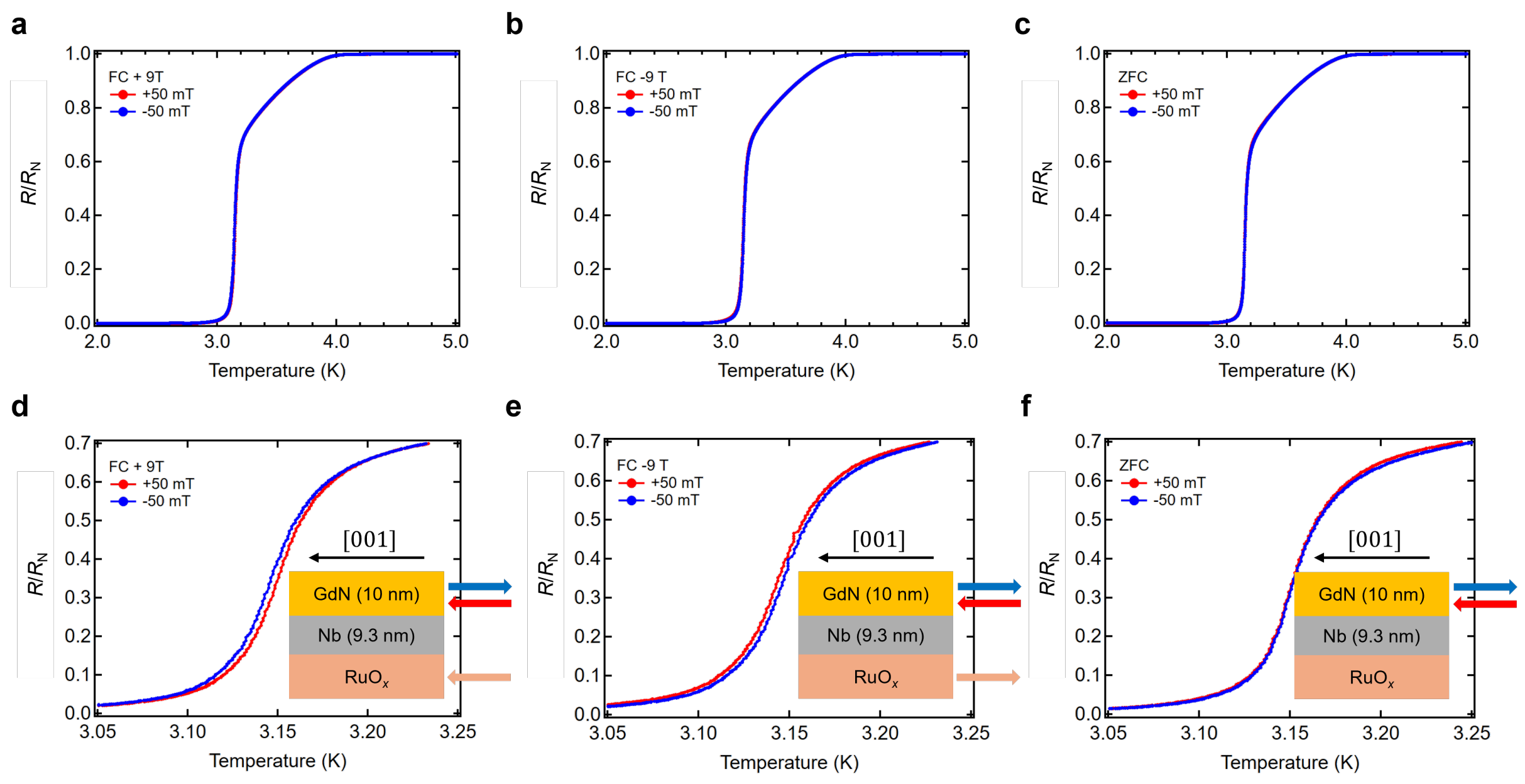

FIG. S3. (a–c) Temperature dependence of the normalized resistance for the AlN(10 nm)/GdN(10 nm)/Nb(9.3 nm)/$RuO_2$(110) (10 nm)//$TiO_2$(110) sample measured with a current applied along the [001] direction under an external magnetic field of ±50 mT after FC at +9 T, FC at −9 T, and ZFC, respectively. (d–f) Enlarged views of the superconducting transition regions in (a–c), respectively. Red and blue solid lines indicate the result of +50 mT and −50 mT, respectively. The accompanying schematics indicate the relative orientation between the exchange field from $RuO_x$ and the magnetisation of GdN.

## S4 In-plane magnetic field dependence of $T_c$ in GdN/Nb

Figure S4 shows the in-plane magnetic field dependence of the superconducting transition temperature $T_c$ of an AlN(10 nm)/GdN(10 nm)/Nb(9.3 nm)//$SiO_2$/Si heterostructure. The external magnetic field is parallel to the current. The inset shows an enlarged view of the low-field region. For a thin superconducting film under an in-plane magnetic field, the suppression of $T_c$ is approximated within the two-dimensional Ginzburg–Landau framework as [S3],

$$T_c(H) = T_c(0)\left[1 - \left(\frac{H}{H_{c2}(0)}\right)^2\right]. \tag{S1}$$

In a superconductor/ferromagnetic-insulator bilayer, an additional exchange field can be induced in the superconductor through the magnetic proximity effect. When this proximity-induced exchange field is parallel to the Zeeman field generated by the applied magnetic field, the total spin splitting in the superconductor increases with increasing magnetic field, leading to a monotonic suppression of $T_c$. In contrast, when the induced exchange field is antiparallel to the applied-field contribution, the two fields can partially compensate each other. Such exchange-field compensation can result in a nonmonotonic $T_c(H)$ dependence and, in the appropriate regime, reentrant superconducting behavior [S4, 5].

As shown in Fig. S4, $T_c$ decreases monotonically with increasing magnetic field. This behavior is therefore consistent with the exchange field induced by GdN in Nb being aligned such that it enhances, rather than compensates. A small cusp is observed in the low-field region, as highlighted in the inset. We tentatively attribute this feature to a small component of the GdN magnetisation that does not fully follow the external magnetic field during magnetisation reversal.Note that, because the superconducting transition becomes broader with increasing magnetic field, $T_c$ was defined at $R = 0.25R_N$.

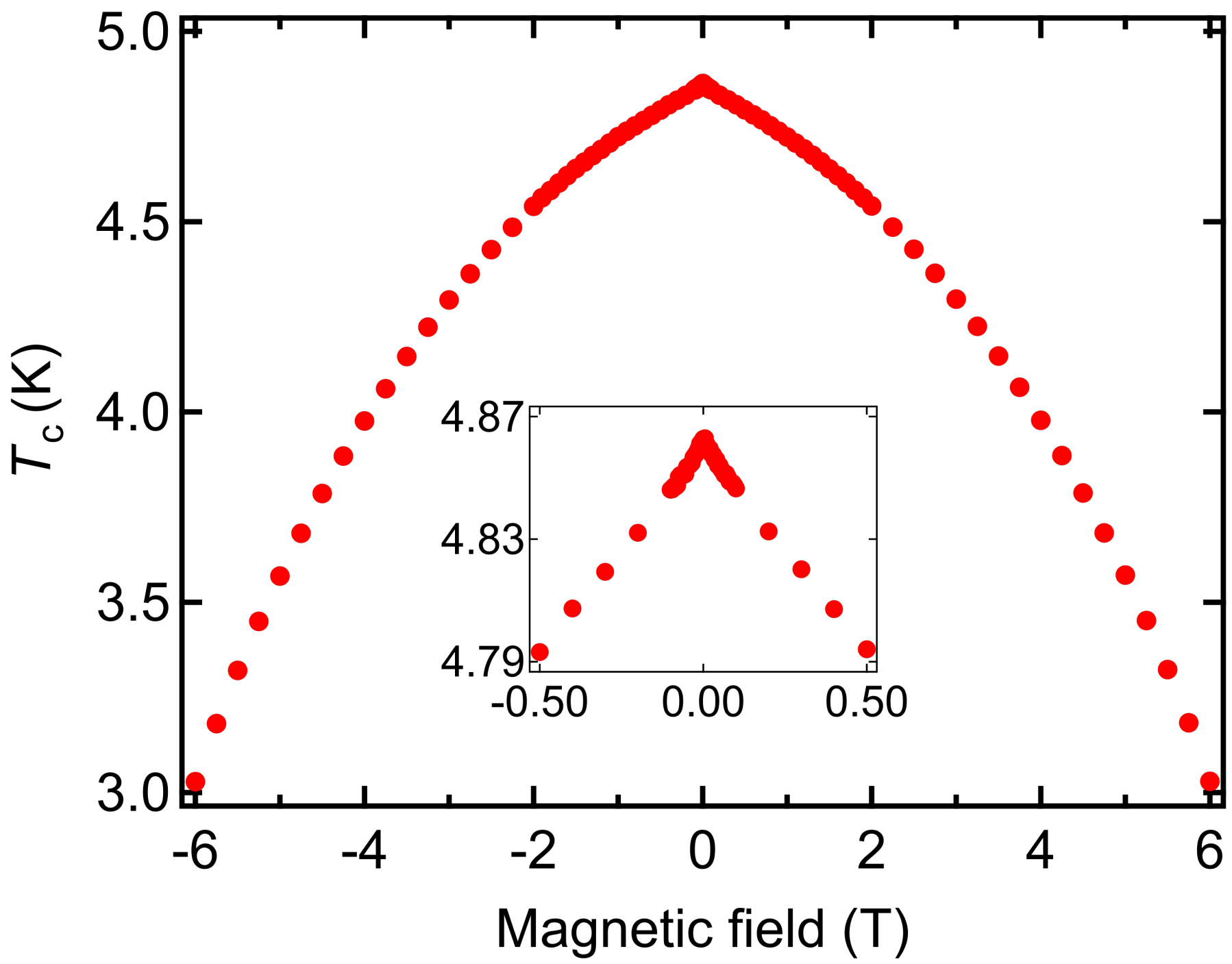

FIG. S4. Magnetic field dependence of $T_c$ of AlN(10 nm)/GdN(10 nm)/Nb(9.3 nm)//$SiO_2$/Si heterostructures. The inset shows an enlarged view of the low-field region.

## S5 Crystallographic-orientation dependence of the superconducting switch

Figures S5a and b show the results obtained after FC along the $[1\bar{1}0]$ direction, with the subsequent magnetic field, and hence the GdN magnetisation, also aligned along the same direction. Figures S5c and d show the corresponding measurements for field cooling and magnetic-field application along the 45° direction, which is defined here as the in-plane direction halfway between [001] and $[1\bar{1}0]$. For both orientations, a clear modulation of $T_c$ is observed, comparable to that obtained for the [001] direction.

Within the proposed altermagnetic picture of $RuO_2$, the Néel vector has been reported to preferentially align along the crystallographic [001] axis, and recent first-principles calculations predict a strong magnetocrystalline anisotropy stabilizing this orientation [S6, 7]. If the observed $\Delta T_c$ originated predominantly from such an altermagnetic order, one would therefore expect the effect to depend strongly on the field orientation and to be substantially reduced when the field is applied away from the preferred Néel-vector direction. Instead, a comparable $\Delta T_c$ is observed for the [001], $[1\bar{1}0]$, and 45° directions. This weak angular dependence disfavors an interpretation based on the conventional altermagnetic order of $RuO_2$ and instead suggests that the $RuO_2$-side magnetic state relevant to the superconducting switching is nearly isotropic with respect to the in-plane field direction.

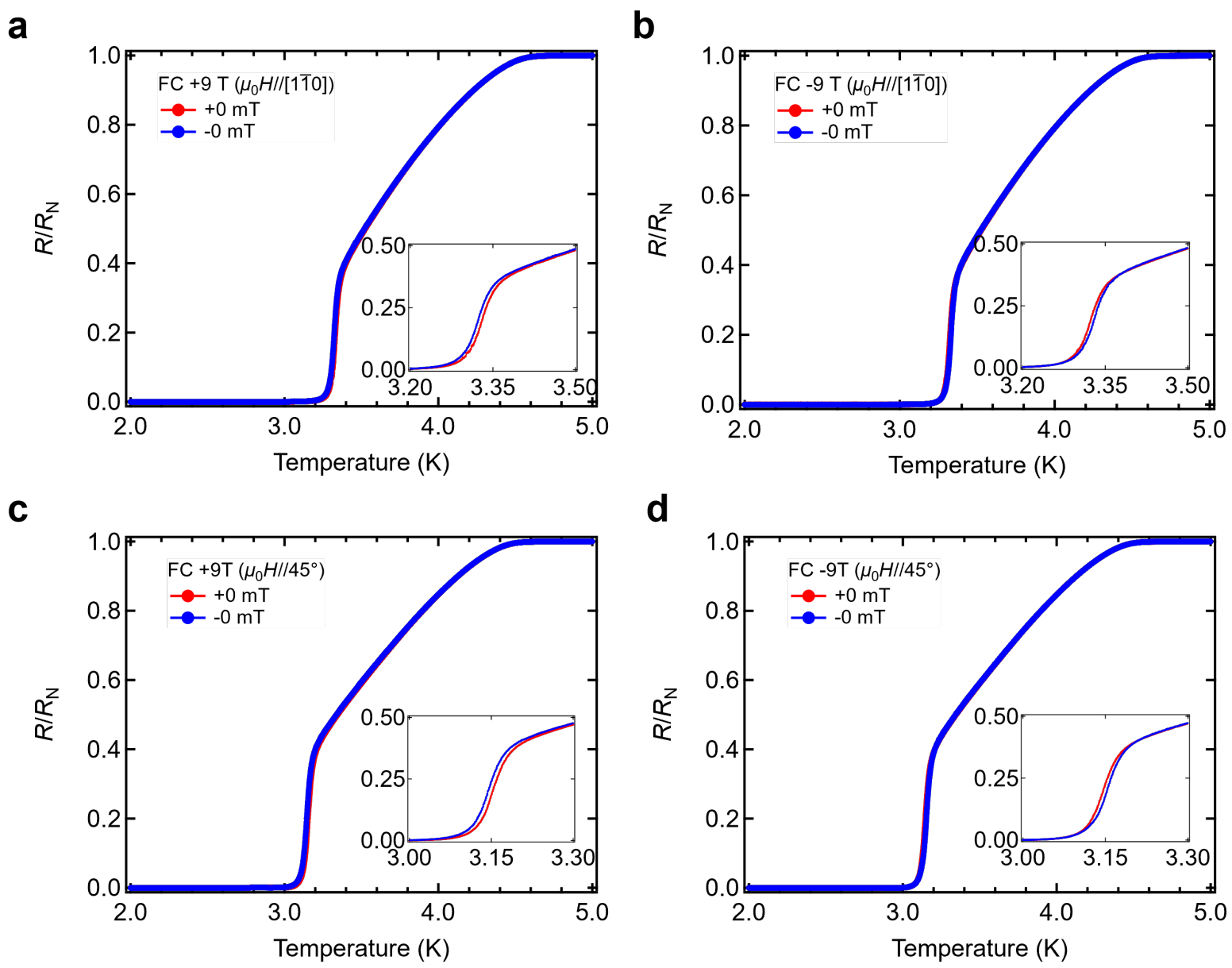

FIG. S5. (a,b) Temperature dependence of the normalized resistance measured after FC ±9 T along the $[1\bar{1}0]$ direction, with the subsequent magnetic field, respectively. (c,d) Corresponding measurements for FC ±9 T and magnetic-field application along the 45° direction, respectively. Red and blue solid lines represent the measurements at +0 and −0 mT, respectively. The insets show enlarged views of the superconducting transitions.

**S6 Temperature dependence of the normalized Nb resistance in the orthogonal magnetic configuration**

In a superconducting switch, no difference in $T_c$ is expected between the relative magnetic configurations of 90° and 270°. To examine this situation, we measured $T_c$ of the 9.3-nm-thick Nb sample in the orthogonal magnetic configuration. Figure S6a shows the device structure and measurement configuration. To avoid possible geometrical effects associated with applying the magnetic field along the short-axis direction of the wire, we used a device geometry rotated by 90° with respect to the device shown in Fig. 1d of the main text. For the FC process, a magnetic field of ±9 T was applied along the [001] direction. Figure S6b schematically illustrates the relative configuration between the $RuO_x$ exchange field and the GdN magnetisation. Figures S6c–f show the temperature dependence of the normalized resistance measured under magnetic fields applied along the $[1\bar{1}0]$ direction. For both ±0 and ±50 mT, almost no difference in $T_c$ is observed between the two magnetic-field directions as shown in Fig. S6c-f. These results are consistent with the superconducting switch model, in which $T_c$ is unchanged for orthogonal magnetic configurations. The small residual difference is likely attributable to a slight misalignment of the magnetisation direction and the device geometry. Note that under the ±0 mT condition, the difference of $T_c$ is sensitive to the magnetic-field angle, and a finite difference can appear even for a small angular deviation from the ideal orthogonal configuration. This sensitivity may be related to the difference between the saturation and remanent magnetisation states of GdN, although further investigation is required to clarify its microscopic origin.

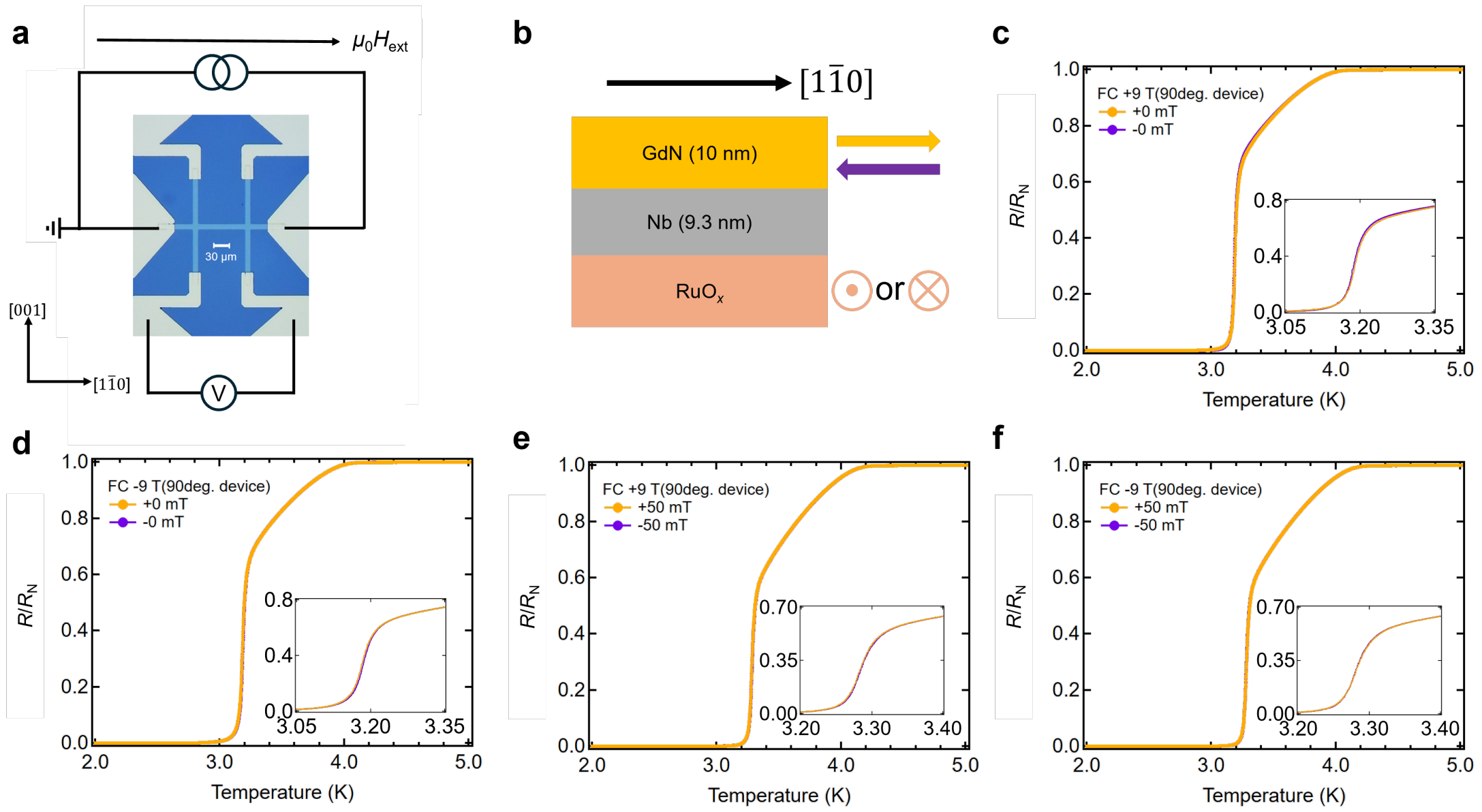

FIG. S6. (a) Device geometry rotated by 90° with respect to the device shown in Fig. 1d of the main text and the measurement configuration. (b) Relative configuration between the $RuO_x$ exchange field and the GdN magnetisation. (c,d) Normalized resistance as a function of temperature for the 9.3-nm-thick Nb sample measured under ±0 mT after FC at +9 T and −9 T, respectively. The yellow and purple solid lines represent the data measured at +0 mT and −0 mT, respectively. The insets show enlarged views near the superconducting transition temperature. (e,f) Normalized resistance of the 9.3-nm-thick Nb sample as a function of temperature measured under ±50 mT after FC at +9 T and −9 T, respectively. The yellow and purple solid lines represent the data measured at +50 mT and −50 mT, respectively. The insets show enlarged views near the superconducting transition temperature.

## S7 Superconducting coherence length of Nb

Figure S7a shows the out-of-plane magnetic-field dependence of the normalized resistance for the AlN(10 nm)/Nb(9.3 nm)/$RuO_2$(10 nm)//$TiO_2$(110) sample measured at various temperatures. Here, the upper critical field, $\mu_0 H_{c2}$, was defined as the external magnetic field at which the normalized Nb resistance is 0.5. Figure S7b shows the temperature dependence of $\mu_0 H_{c2}$ on the left axis and the temperature dependence of the normalized resistance at zero magnetic field on the right axis. The fitting was performed in the temperature range close to $T_c$ using the relation [S3, 8]

$$\mu_0 H_{c2}(T) = \mu_0 H_{c2}(0)(1 - T/T_c). \tag{S2}$$

The superconducting coherence length, $\xi$, was then estimated from

$$\xi = \sqrt{\frac{\Phi_0}{2\pi\mu_0 H_{c2}(0)}}, \tag{S3}$$

where $\Phi_0$ is the quantum superconducting flux [S9]. From the fitting, $\mu_0 H_{c2}(0)$ was estimated to be 2.56 $\pm$ 0.008 T, yielding a superconducting coherence length ($\xi$) of 11.35 $\pm$ 0.02 nm. These results indicate that the 9.3-nm-thick Nb sample used in the main text is thinner than $\xi$, whereas the 12-nm-thick Nb sample in S8 is thicker than $\xi$.

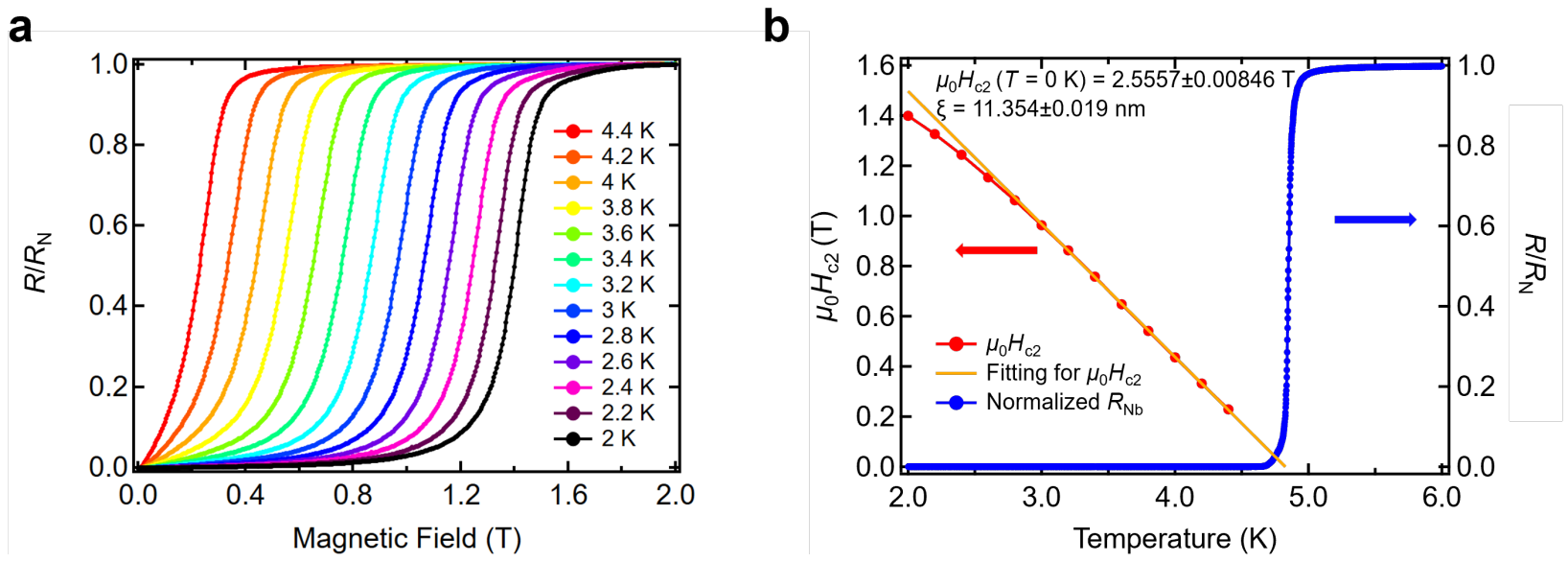

FIG. S7. (a) Out-of-plane magnetic-field dependence of the normalized resistance for the AlN(10 nm)/Nb(9.3 nm)/$RuO_2$(110) (10 nm)//$TiO_2$(110) sample measured at various temperatures. (b) Temperature dependence of $\mu_0 H_{c2}$ (left axis and red dots) and normalized resistance at zero magnetic field (right axis and blue dots). The orange solid line represents the fitting result for the temperature dependence of $\mu_0 H_{c2}$.

## S8 Superconducting switching for an Nb thickness exceeding the coherence length

In the superconducting switch, when the superconducting layer thickness exceeds $\xi$, Cooper pairs cannot effectively experience the exchange fields from both the upper and lower magnetic layers simultaneously. Therefore, the difference of $T_c$ originating from the relative magnetic configuration is significantly reduced compared with the case where the superconducting layer thickness is smaller than $\xi$ [S10]. Figures S8a–d show the temperature dependence of the normalized resistance for the AlN(10 nm)/GdN(10 nm)/Nb(12 nm)/$RuO_2$(110) (10 nm)//$TiO_2$(110) sample. Both the FC field and the magnetic field applied were oriented along the [001] direction. The Nb thickness of 12 nm is thicker than the estimated $\xi$ in S7. Consistently, $T_c$ shift depending on the relative configuration between the $RuO_x$ exchange field and the GdN magnetisation is strongly suppressed under both the $\pm 0$ mT conditions and the $\pm 50$ mT conditions as shown in Fig. S8. These results further support that the $T_c$ difference observed in Fig. 2 of the main text is not caused by extrinsic effects such as a residual magnetic field or a stray field of GdN but rather originates from the exchange field across the Nb layer.

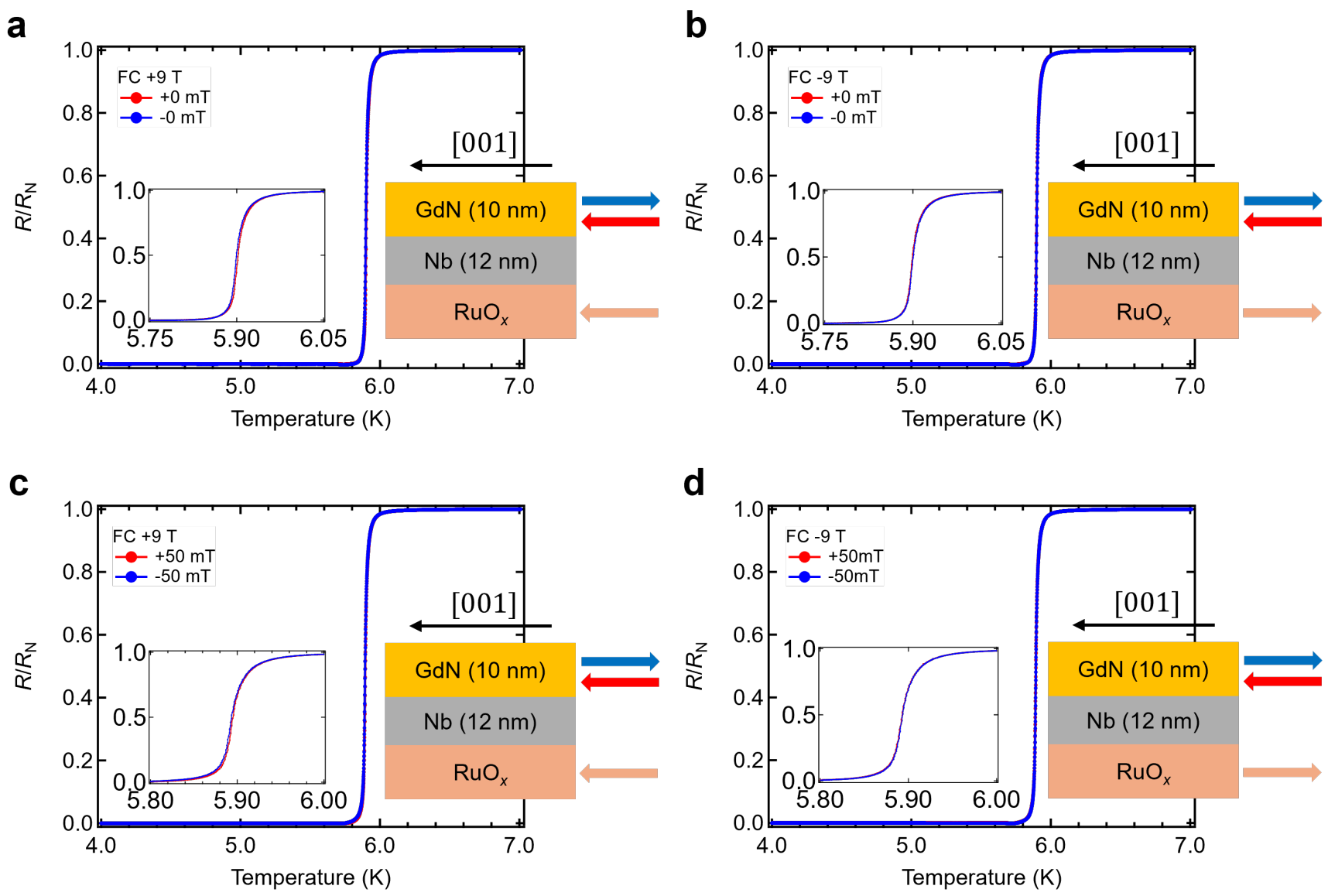

FIG. S8. (a–d) Normalized Nb resistance as a function of temperature for the 12-nm-thick Nb sample. The measurements were performed under ±0 mT in (a,b) and under ±50 mT in (c,d). The red and blue solid lines represent the data measured at +0 mT and −0 mT in (a,b), and at +50 mT and −50 mT in (c,d), respectively. The insets show enlarged views near the superconducting transition temperature. The schematic illustrations indicate the relative configurations of the exchange fields.

## S9 Temperature dependence of the normalized Nb resistance in control samples

The results presented in the main text originate from the combined effect of the $RuO_x$ exchange field and the GdN exchange field; therefore, a similar $T_c$ difference is not expected when either contribution is absent.

Figures S9a and b show the temperature dependence of the normalized resistance for the AlN(10 nm)/GdN(10 nm)/Nb(9.3 nm)//$SiO_2$/Si control sample after FC at +9 T and −9 T, respectively. In both FC conditions, no clear change in $T_c$ depending on the GdN magnetisation direction is observed. This control experiment allows us to exclude a possible superconducting switching mechanism arising from additional magnetic regions at the GdN/Nb interface. In particular, structural disorder or amorphous patches at the GdN/Nb interface could, in principle, host a magnetic state distinct from the bulk GdN magnetisation, resulting in an effective SC/FM(FerroMagnet)/FM-type structure. If the relative orientation between such interfacial magnetism and the GdN magnetisation produced a superconducting spin-valve effect, reversing the FC direction would be expected to induce a measurable change in $T_c$. The absence of a clear $T_c$ difference between the +9 T and −9 T FC conditions indicates that this scenario does not account for the superconducting switching observed in the main sample.

Figures S9c and d show the temperature dependence of the normalized resistance for the AlN(10 nm)/Nb(9.3 nm)/$RuO_2$(10 nm)//$TiO_2$(110) control sample under ±50 mT after FC at +9 T and −9 T, respectively. In both cases, no clear change in $T_c$ depending on the external magnetic-field direction is observed. In principle, a $T_c$ difference could arise from the relative orientation between the $RuO_x$ exchange field and the external magnetic field in this control sample. However, such a shift was not clearly resolved in the present measurements, suggesting that any $T_c$ difference originating solely from the $RuO_2$ exchange field, if present, is below the experimental detection limit. These control experiments indicate that the coexistence of the $RuO_x$ exchange field and the GdN magnetisation is essential for the $T_c$ difference observed in the main text.

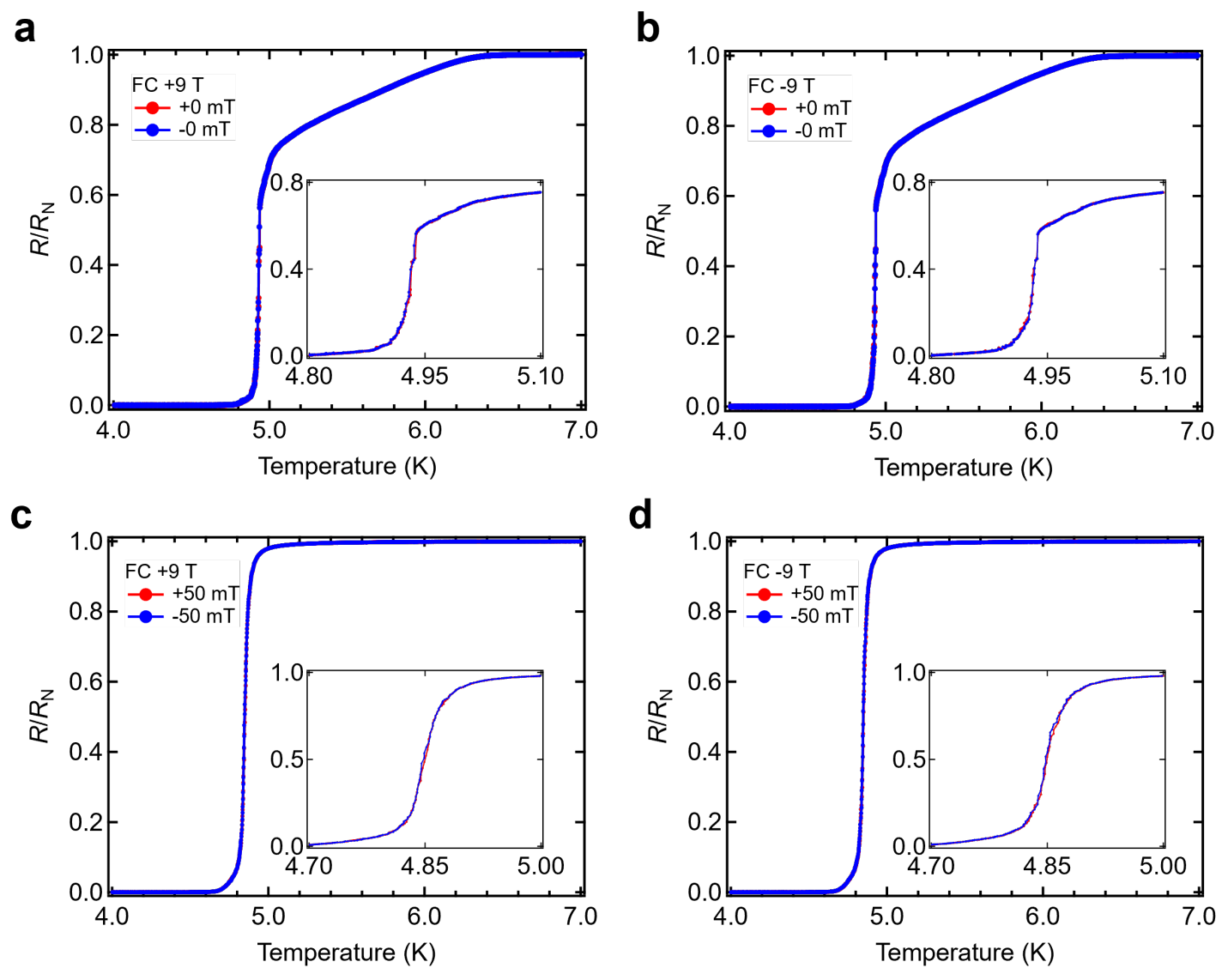

FIG. S9. (a,b) Temperature dependence of the normalized resistance of the AlN(10 nm)/GdN(10 nm)/Nb(9.3 nm)//$SiO_2$/Si control sample after FC at (a) +9 T and (b) −9 T. The red and blue solid lines represent the measurements at +0 mT and −0 mT, respectively. (c,d) Temperature dependence of the normalized resistance of the AlN(10 nm)/Nb(9.3 nm)/$RuO_2$(10 nm)//$TiO_2$(110) control sample measured under ±50 mT after FC at (c) +9 T and (d) −9 T. The red and blue solid lines represent the measurements under +50 mT and −50 mT, respectively.

## S10 Superconducting switching in the 50-nm-thick $RuO_2$ sample

$RuO_2$ is known to exhibit metallic transport except in the ultrathin limit [S11]. Therefore, in the present experiment, the possibility that Cooper pairs penetrate into metallic $RuO_2$ cannot be excluded. To rule out the possibility that a small magnetic moment at the $TiO_2/RuO_2$ interface contributes to the superconducting switch effect, we performed superconducting switch measurements using an AlN(10 nm)/GdN(10 nm)/Nb(9.3 nm)/$RuO_2$(50 nm)//$TiO_2$ heterostructure. For a $RuO_2$ thickness of 50 nm, considering the coherence length estimated in Fig. S8, the Cooper pair is not expected to extend as far as the $TiO_2/RuO_2$ interface. The measurements were performed at a current density of $4.30 \times 10^8$ A/m$^2$.

The sample was first field-cooled under a magnetic field of 9 T, after which the magnetic state on the $RuO_2$ side was controlled by applying saturation fields of $\pm1.5$ T. Figures S10a and b show the temperature dependence of the normalized resistance measured at $\pm0$ mT after saturation at $\pm1.5$ T. Although $T_c$ is reduced compared with the 10-nm-thick $RuO_2$ sample, likely due to the increased $RuO_2$ thickness, a clear difference in $T_c$ depending on the saturation-field direction and the magnetisation direction of GdN is still observed, similar to the case of 10-nm-thick $RuO_2$. Importantly, the polarity of the superconducting switching is also the same.

These results indicate that the superconducting switch effect discussed in this work originates from magnetism at the $RuO_2$/Nb interface rather than from magnetism at the $TiO_2/RuO_2$ interface.

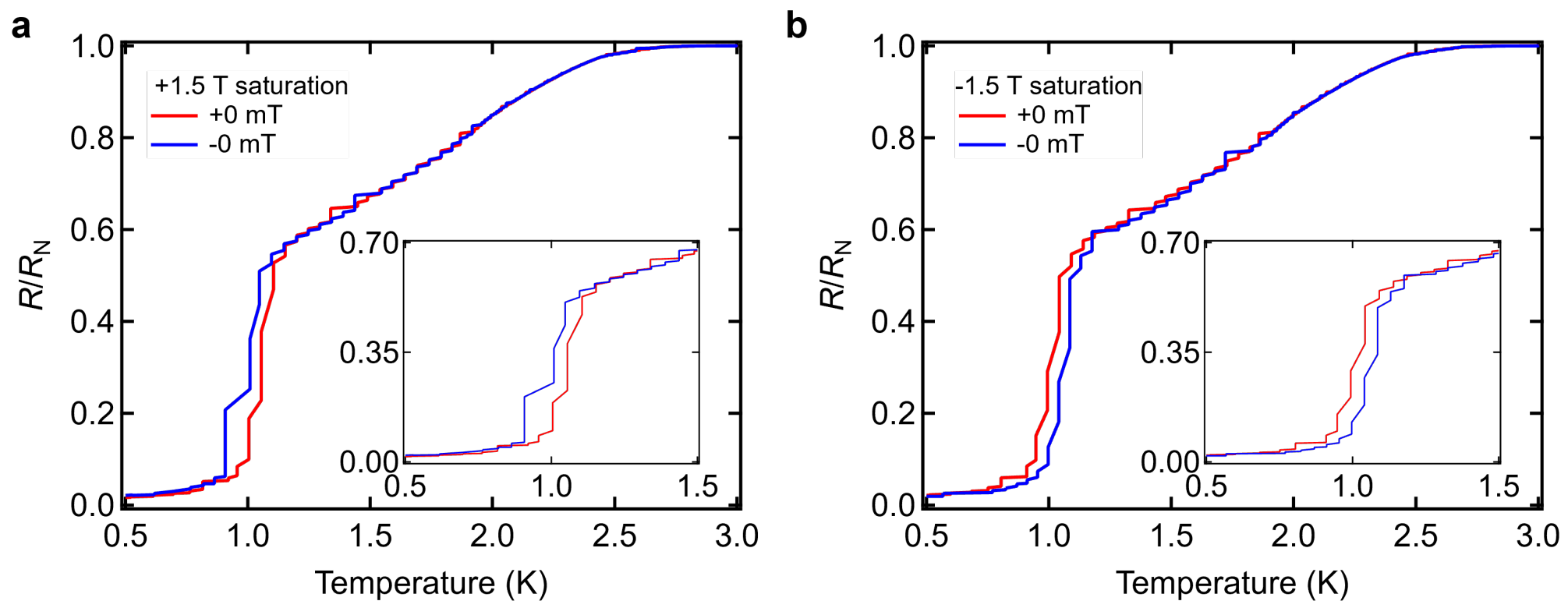

FIG. S10. (a,b) Temperature dependence of the normalized resistance of the AlN(10 nm)/GdN(10 nm)/Nb(9.3 nm)/$RuO_2$(50 nm)//$TiO_2$ heterostructure measured after saturation at ±1.5 T, respectively. The red and blue solid lines represent the measurements at +0 and −0 mT, respectively. The insets show enlarged views of the superconducting transitions.

## S11 Origin of the magnetism in the $RuO_x$ redox layer

Reduction of $Ru^{4+}$ at the Nb/$RuO_2$ interface produces $Ru^{3+}$, although further local reduction to $Ru^{2+}$ may also occur. However, oxygen vacancies in $RuO_2$ are known to induce electron accumulation at neighboring Ru sites and stabilize substantial fractions of low-valent $Ru^{3+}$ species[S12, 13]. XPS studies have directly resolved enhanced $Ru^{3+}$ components in oxygen-deficient $RuO_2$, while Ru K-edge XANES measurements yield average Ru valences as low as approximately +3.4, consistently interpreted as a mixed $Ru^{4+}/Ru^{3+}$ state [S12–14]. Although a fraction of the interfacial Ru sites may be further reduced to lower-valence states, a substantial population of $Ru^{3+}$ sites is nevertheless expected to persist. For the magnetic physics considered here, the relevant reduced species is $Ru^{3+}$, which carries one hole in the $t_{2g}$ multiplet ($t_{2g}^5$), forming in the atomic limit a Kramers doublet with a robust $S = 1/2$ moment, in contrast to $Ru^{4+}$, whose moment formation relies on the weak Hund's coupling and is quenched by hybridization in stoichiometric $RuO_2$.

The reduction of one quarter of the oxygens is not a dilute perturbation: each oxygen is shared by three Ru atoms, so 25% oxygen deficiency leaves on average 1.5 missing ligands per $RuO_6$ octahedron, and the loss of a single ligand removes an antibonding shift of order 0.3–0.5 eV from the $t_{2g}$ orbital pointing along the broken bond—comparable to the bandwidth. Organizing the problem by symmetry rather than amplitude, one finds a conclusion independent of the octahedron orientation and of the $RuO_2$ polymorph: at each reduced site the orbital doublet with lobes along the local vacancy axis $\hat{\mathbf{n}}_v$ is lowered, and the magnetically active hole resides in the remaining planar orbital, whose band is narrowed by the missing ligands. Two consequences follow. First, the Stoner criterion $U\nu_F \gtrsim 1$, far from satisfied in stoichiometric $RuO_2$, becomes attainable in the redox layer: the same vacancies that create the $Ru^{3+}$ holes carrying the magnetic moment also narrow the band in which they order. The order is ferromagnetic because the layer is metallic and mixed-valent, while structural disorder frustrates any coherent antiferromagnetic pattern. Second, through the spin-orbit coupling each vacancy axis becomes a local magnetic hard axis: a single-site anisotropy energy $K(\mathbf{m} \cdot \hat{\mathbf{n}}_v)^2$ with $K > 0$, where $\mathbf{m}(\mathbf{r})$ is the coarse-grained Stoner order parameter of the layer.

The vacancy axes are random in direction, but not isotropically so: oxygen leaves through the Nb interface, so the distribution of $\hat{\mathbf{n}}_v$ is biased toward the normal $\hat{\mathbf{e}}_z$, with a quadrupolar alignment parameter $P_v = \langle P_2(\hat{\mathbf{n}}_v \cdot \hat{\mathbf{e}}_z)\rangle > 0$. The anisotropy energy of the layer then splits into a coherent part, $\bar{K} P_v m_z^2$—an easy plane, since the mean hard axis is the normal—and a zero-mean random-axis part. Together with the exchange stiffness of the itinerant magnet, this makes the layer a random-anisotropy magnet of the type studied in amorphous systems [S15], with additional Dzyaloshinskii–Moriya and Rashba-type couplings generated by the parity-odd part of the vacancy fields.

This picture is consistent with the phenomenology of $RuO_x$ magnetisation. In such a magnet, the exchange stiffness favours uniform order, whereas the random anisotropy favours local alignment with the random axes. As argued by Imry and Ma [S16, 17], the energy gained by following the random axes over a region of size $L$ grows faster with $L$ than the exchange cost of rotating the magnetisation, in any dimension below four. Long-range order is therefore expected to be destroyed beyond a characteristic length, the Imry–Ma length $L^*$, set by the exchange stiffness and by the strength and correlation length of the vacancy disorder, which are not determined here; the layer would then break into nearly collinear in-plane domains whose orientation wanders beyond $L^*$. The total moment of a nanometer-thick layer is small, and if the layer breaks into such domains, their partial alignment by field cooling would reduce the domain-averaged moment $\bar{\mathbf{m}}$ further, so that it falls below standard magnetometry resolution: the magnet is fragile in moment, and $M(H = 0) \approx 0$ is expected rather than paradoxical. Its rigidity follows from the same smallness: an external field couples to the layer only through a small Zeeman energy, so that reversal—whether by coherent rotation or by depinning of domain walls from the random-anisotropy landscape [S18]—requires large fields even for a moderate anisotropy energy. The same oxygen vacancies that enable the Stoner instability (band narrowing) also pin its order parameter (random axes). A fragile yet rigid magnet is thus a natural outcome of interfacial redox chemistry.

Two further properties matter for the switch. First, the easy plane is a gradient effect, strongest for thin redox layers; a thick redox layer would approach an isotropic random-axis glass with no switchable moment, acting on the superconductor only as a spin-disorder pair breaker. The observation of a hysteretic, switchable subsystem is thus consistent with the thin, self-limited redox depth. Second, the layer is in-plane isotropic—$\bar{\mathbf{m}}$ is slaved to the cooling field, not to crystal axes. This distinguishes the redox scenario sharply from any altermagnetic or surface-crystalline magnetism, which would imprint the symmetry of the

(110) surface on the angular response, and it is consistent with the observed insensitivity of the switch amplitude to the FC angle in the plane (see S6 for the 45° data for in-plane anisotropy).

The electronic structure of the reduced layer also accounts for the negative sign of the exchange field inferred in the main text. In the $t_{2g}^5$ configuration of $Ru^{3+}$ the $t_{2g}$ band is more than half filled: the majority-spin $t_{2g}$ states are occupied, and the only empty $t_{2g}$ states near the Fermi level are minority-spin states. Hybridization of Nb conduction electrons with these empty states lowers the energy of minority-spin Nb electrons, inducing a spin polarization in Nb antiparallel to the $RuO_x$ magnetisation, i.e. an antiferromagnetic coupling between Nb conduction electrons and the Ru moments. The same sign follows in a localized picture (atomic limit), since a Nb electron can virtually hop into the $Ru^{3+}$ hole only if its spin is antiparallel to the local moment. The mechanism yields the opposite sign for GdN, where the lowest empty Gd $5d$ states are majority-spin states owing to their exchange splitting by the $4f^7$ moment, in agreement with the positive exchange field found for the GdN control sample (Supplementary S5).

## S12 Theory II: Quasiclassical theory about $\Delta T_c$

To determine the critical temperature of the diffusive superconducting film, we solve the Usadel equation [S19] subject to the appropriate boundary conditions for superconductors in contact with ferromagnetic insulators [S20].

Near the superconducting transition temperature, the Usadel equation can be linearized. For Nb thicknesses satisfying $d_{\mathrm{Nb}} \lesssim \xi$, the superconducting order parameter can be approximated as uniform across the film thickness. Averaging the linearized equation over the film thickness then yields the following self-consistency relation for $T_c$:

$$\ln \frac{T_{c0}}{T_c} = \frac{(1+\delta)\psi\left(z_+\right) + (1-\delta)\psi\left(z_-\right)}{2} - \psi\left(\frac{1}{2}\right), \tag{S4}$$

where $\psi(z)$ is the digamma function and $T_{c0}$ is the superconducting transition temperature of the reference Nb film in the absence of interfacial magnetic depairing. The parameters $\delta$ and $z_\pm$ are defined as:

$$\delta = \frac{\tau_{sf}^{-1} - \tau_{so}^{-1}}{\sqrt{\left(\tau_{sf}^{-1} - \tau_{so}^{-1}\right)^2 - \left(\frac{4h}{\hbar}\right)^2}} \tag{S5}$$

$$z_\pm = \frac{1}{2} + \frac{2D\vec{q}^{\,2} + 2\tau_{sf}^{-1} + \tau_{so}^{-1} \pm \sqrt{\left(\tau_{sf}^{-1} - \tau_{so}^{-1}\right)^2 - \left(\frac{4h}{\hbar}\right)^2}}{8\pi \frac{k_B}{\hbar} T_c}. \tag{S6}$$

Here, $\tau_{sf}$ denotes the effective spin-relaxation time associated with the ferromagnetic-insulator interfaces. By contrast, $\tau_{so}$ characterizes spin relaxation arising from spin–orbit coupling in Nb. The parameter $h$ is the effective exchange field averaged over the superconducting layer, induced by the two magnetic interfaces. For the parallel (P) and antiparallel (AP) magnetic configurations, the effective exchange fields are:

$$h_{P/AP} = h_{\mathrm{GdN}} \pm h_{\mathrm{RuO_x}}. \tag{S7}$$

The gauge-invariant superfluid wavevector is defined as $\vec{q} = \vec{\nabla}\varphi - \frac{2e}{\hbar}\vec{A}$ where $\varphi$ is the superconducting phase and $\vec{A}$ is the vector potential. In the limit $q \to 0$, corresponding to the absence of depairing associated with finite condensate momentum, Eq. (S4) formally resembles the expression describing the suppression of superconductivity by magnetic impurities [S21]. The underlying physical mechanisms are nevertheless distinct. Here, the effective exchange field and spin-relaxation processes arise from the spin-mixing conductances at the superconductor/ferromagnetic-insulator interfaces, rather than from a bulk magnetic field or randomly distributed magnetic impurities. Accordingly, $h$ and $\tau_{sf}$ should be interpreted as effective parameters characterizing the interfacial exchange and spin-relaxation effects, respectively.

For Nb, we take the spin–orbit relaxation time to be $\tau_{so} \approx 0.2$ ps, as estimated from Ref. [S22]. To estimate the effective exchange field induced by GdN, we use the experimental results reported in Ref. [S23], which investigated a symmetric GdN/Nb/GdN spin valve. For a Nb thickness of $d_{\mathrm{Nb}} = 10$ nm, the reported transition temperatures were $T_{c0} \approx 6.7$ K for the bare film and $T_c^{AP} \approx 3.9$ K and $T_c^{P} \approx 3.7$ K for the valve. Using these values in Eq. (S4), we obtain the effective spin-relaxation time and exchange field:

$$\tau_{sf} \approx 1.74 \text{ ps}, \quad h_{\mathrm{GdN}} \approx 0.12 \text{ meV}. \tag{S8}$$

The relation between the thickness-averaged field $h_{\mathrm{GdN}}$ and the interfacial exchange $J_{\mathrm{GdN}}$ is [S24]:

$$J_{\mathrm{GdN}} = \frac{d_{\mathrm{Nb}}}{a_{\mathrm{Nb}}} h_{\mathrm{GdN}} \approx 3.63 \text{ meV} \tag{S9}$$

where $a_{\mathrm{Nb}} = 0.33$ nm is the lattice parameter for Nb.

For the GdN/Nb/$\mathrm{RuO_x}$ spin valve investigated here, the Nb thickness is $d_{\mathrm{Nb}} = 9.3$ nm. Assuming that the interfacial exchange field $J_{\mathrm{GdN}}$ is unchanged, we obtain the corresponding thickness-averaged contribution:

$$h_{\mathrm{GdN}} = \frac{a_{\mathrm{Nb}}}{d_{\mathrm{Nb}}} J_{\mathrm{GdN}} \approx 0.13 \text{ meV}. \tag{S10}$$

We define $T_{c0}$ using a reference AlN(10 nm)/Nb(9.3 nm)/$\mathrm{SiO_2}$/Si sample. As shown in Fig. S12, its superconducting transition temperature is $T_{c0} = 6.00$ K.

Under an applied current density of $5.38 \times 10^8$ A/m$^2$, the measured transition temperatures of our spin valve are $T_c^{AP} \approx 3.186$ K and $T_c^{P} \approx 3.194$ K. Using these values in Eq. (S4), together with the estimated GdN contribution, we obtain:

$$\tau_{sf} \approx 1.8 \text{ ps}, \quad h_{\mathrm{RuO_x}} \approx -0.004 \text{ meV}. \tag{S11}$$

Finally, using the same thickness-scaling relation as in Eq. (S9), we estimate the interfacial exchange field associated with the $\mathrm{RuO}_x$ layer:

$$J_{\mathrm{RuO_x}} = \frac{d_{\mathrm{Nb}}}{a_{\mathrm{Nb}}} h_{\mathrm{RuO_x}} \approx -0.11 \text{ meV}. \tag{S12}$$

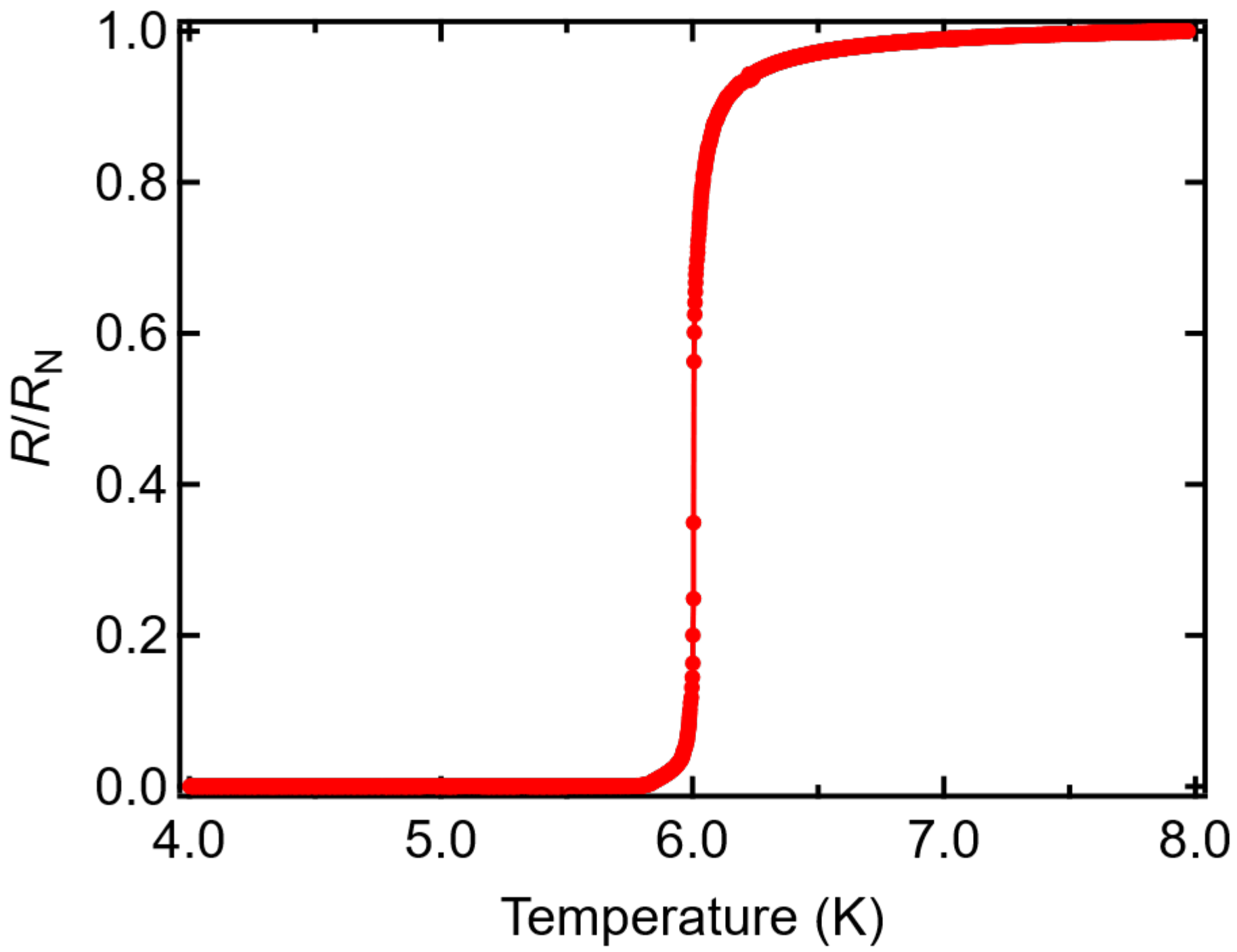

FIG. S12. Superconducting transition temperature of the AlN(10 nm)/Nb(9.3 nm)//SiO2/Si sample.